\documentclass[prd,aps,nofootinbib,onecolumn,showpacs,10pt]{revtex4}
\usepackage{graphicx,epsfig}

\def\dd{\langle \bar d d \rangle}

\def\qq{\langle \bar q q \rangle}

\newcommand {\ttn}[1]{\times10^{#1}}
\newcommand{\err}[2]{^{+ #1}_{- #2}}
\newcommand{\degr}{^\circ}
\newcommand{\lpm}{\lp\lm}
\newcommand{\epm}{e^+e^-}
\newcommand{\mupm}{\mu^+\mu^-}
\newcommand{\taupm}{\tau^+\tau^-}
\newcommand{\eqn}[2]{\begin{equation}\label{#1}#2\end{equation}}
\def  \to     {\rightarrow}
\def  \bcen   {\begin{center}}
\def  \ecen   {\end{center}}
\def  \beq    {\begin{equation}}
\def  \eeq    {\end{equation}}
\def  \beqa   {\begin{eqnarray}}
\def  \eeqa   {\end{eqnarray}}
\def  \nn     {\nonumber}
\def  \g      {\gamma}

\def  \lm        {\ell^-}
\def  \lp        {\ell^+}

\def  \kkka        {K_{1A}}
\def  \kkkb        {K_{1B}}
\def  \kkkab        {K_{1(A,B)}}
\def  \kkkl        {K_{1}(1270)}
\def  \kkkh        {K_{1}(1400)}
\def  \kkklh        {K_{1}(1270,1400)}
\def  \toka         {B \to K_{1A} \lp \lm}
\def  \tokb         {B \to K_{1B} \lp \lm}
\def  \tok         {B \to K_1 \lp \lm}
\def  \tokl         {B \to K_{1}(1270) \lp \lm}
\def  \tokh         {B \to K_{1}(1400) \lp \lm}
\def  \tokab         {B \to K_{1(A,B)} \lp \lm}
\def  \toklh         {B \to K_{1}(1270,1400) \lp \lm}
\def  \Mtokl          {\bra{\kkkl}J_\mu\ket{B}}
\def  \Mtokh        {\bra{\kkkh}J_\mu\ket{B}}
\def  \Mtoklh     {\bra{\kkklh}J_\mu\ket{B}}
\def  \Mtoka     {\bra{\kkka}J_\mu\ket{B}}
\def  \Mtokb       {\bra{\kkkb}J_\mu\ket{B}}
\def  \Mtokab   {\bra{\kkkab}J_\mu\ket{B}}
\def \Bpi {\hat{\Pi}}

\def \bb {\bar{b}}
\def \vrr {\varrho}
\def \corone{\Pi^{A,a}_{\mu\nu}(p^2,p^{\prime2})}
\def \coroneB{\hat{\Pi}^{A,a}_{\mu\nu}(p^2,p^{\prime2})}
\def \cortwo{\Pi^{T,a}_{\mu\nu\rho}(p^2,p^{\prime2})}
\def \cortwoB{\hat{\Pi}^{T,a}_{\mu\nu\rho}(p^2,p^{\prime2})}
\def \ppeps {(p',\epsilon)}
\def \ppsq  {p^{\prime 2}}

\newcommand {\epss}[4] {\varepsilon_{#1 #2 #3 #4}}
\newcommand {\ppu}[1]{p^{\prime #1}}
\newcommand {\rru}[1]{r^{\prime #1}}
\def \ffl {f_i^{1270}}
\def \ffh {f_i^{1400}}
\def \ffa {f_{i,A}}
\def \ffb {f_{i,B}}
\def  \fvb     {V_{1}}
\def  \fvi     {V_{2}}
\def  \fvu     {V_{3}}
\def   \ftb  {T_{1}}

\def   \ftu  {T_{3}}

\def  \faa     {A_A}
\def  \fab     {A_B}
\def  \fvba     {V_{1A}}
\def  \fvbb     {V_{1B}}
\def  \fvia     {V_{2A}}
\def  \fvib     {V_{2B}}
\def  \fvua     {V_{3A}}
\def  \fvub     {V_{3B}}
\def   \ftba  {T_{1A}}
\def   \ftbb  {T_{1B}}
\def   \ftia  {T_{2A}}
\def   \ftib  {T_{2B}}
\def   \ftua  {T_{3A}}
\def   \ftub  {T_{3B}}

\newcommand{\onepone}{{1^1P_1}}
\newcommand{\sanpone}{{1^3P_1}}
\newcommand{\konea}{K_{1A}}
\newcommand{\koneb}{K_{1B}}
\newcommand{\bra}[1]{\langle {#1}|}
\newcommand{\ket}[1]{|{#1} \rangle}
\newcommand{\tetam}{{\cal M}_\theta }
\newcommand{\thetaK}{\theta_{K_1}}
\def  \qsq      {q^2}

\newcommand\iden{\leavevmode\hbox{\small1\normalsize\kern-.33em1}}
\def \nn {\nonumber}

\begin{document}
\vspace*{2cm}

\title[$\toklh$ in QCDSR]{THE SEMILEPTONIC $B$ TO $K_1(1270,1400)$ DECAYS IN QCD SUM RULES }

\author{H. Da{\u g}$^{1,2,*}$, A. {\" O}zpineci$^{1,\dagger}$, M. T. Zeyrek$^{1,\ddagger}$}
\address{ \vspace*{0.1in} $^1$ Department of Physics, Middle East
Technical University,\\06531 Ankara, Turkey.\\
$^2$ Laboratory for Fundamental Research , {\"O}zye{\u g}in University,\\
Kusbakisi Cad. No:2, Altunizade-Uskudar,\\ 34662, Istanbul, Turkey.\\
e-mails:~~$^*$huseyin.dag@cern.ch,\\
$^\dagger$ozpineci@metu.edu.tr, $^\ddagger$zeyrek@metu.edu.tr}

\vspace*{1.0cm}

\begin{abstract}
We analyze the semileptonic rare decays of $B$ meson to $K_{1}
(1270)$ and $K_{1} (1400)$ axial vector mesons. The $B\rightarrow
K_{1} (1270,1400) \ell^+ \ell^-$ decays are significant flavor
changing neutral current decays of the $B$ meson. These decays are
sensitive to the new physics beyond SM, since these processes are
forbidden at tree level at SM. These decays occurring at the quark
level via $b\rightarrow s \ell^+ \ell^- $ transition, also provide
new opportunities for calculating the CKM matrix elements $V_{bt}$
and $V_{ts}$. In this study, the transition form factors of the
$B\rightarrow K_{1} (1270,1400) \ell^+ \ell^-$ decays are calculated
using three-point QCD sum rules approach. The resulting form factors
are used to estimate the branching
fractions of these decays.\\

\end{abstract}
 \maketitle



%


\section{Introduction}
In standard model (SM), the flavor changing neutral current (FCNC)
decays of $B$ meson are forbidden at tree level and occur only at
loop level. These decays are good candidates for searching new
physics (NP) beyond SM or the modifications on the SM.  Some of
these rare FCNC decays of B meson; semileptonic and radiative decays
into a vector or an axial vector meson, such as $B \to
K^*(892)\gamma$ \cite{Aubert:2004te,Nakao:2004th,Coan:1999kh}, $B
\to \kkklh \gamma$\cite{Yang:2004as} and $B\to
K^{0*}(892)e^+e^-(\mu^+ \mu^-)$\cite{Ishikawa:2003cp,Aubert:2006vb}
have been observed. For the channel $B\to K^*(892)\lp\lm$, the
measurement of isospin and forward backward asymmetries at BaBar are
also reported\cite{Ishikawa:2006fh,aubert:2008ju,Aubert:2008ps}. For
the radiative decays of B meson into $\kkklh$ axial vector meson
states, Belle reported the following branching
fractions\cite{BelleLH}:
\begin{eqnarray}\label{belleLH}
\nn {\cal B}(B^+\to \kkkl^+ \gamma)&=&(4.28\pm0.94\pm0.43)\times10^{-5},\\
{\cal B}(B^+\to \kkkh^+ \gamma)&<&1.44 \times 10^{-5}.
\end{eqnarray}

The semileptonic $\toklh$ decays are  significant FCNC decays of $B$
meson, which occur via $b\to s \lm \lp$ transitions at quark level.
The semileptonic decay modes $\toklh$ have not been observed yet,
but are expected to be observed in forthcoming $pp$ and $e^+e^-$
accelerators, such as LHC\cite{LhcB} and SuperB\cite{superB}. In
particular LHCb experiment at the LHC could be the first place to
observe such decays\cite{LhcB2}. Observation of these decays might
also provide new opportunities for calculating $V_{bt}$ and
$V_{ts}$; the elements of the Cabibbo-Kobayashi-Maskawa (CKM)
matrix. Recently, some studies on $\toklh$ decays have been
made\cite{yang87,hatanakayang,gaur1,basihry,vali-kazem,Paracha:2007yx,Ahmed:2008ti,Saddique:2008xj,lee}.
Theoretical	studies	of	the decays into final states containing $\kkklh$ axial vector states is rendered more complicated due to the presence of K1 mixing. The
properties of axial vector states are studied in \cite{yangana}. In
QCD, the real physical axial vector $K_1$ states, $K_1(1270)$ and
$K_1(1400)$ are mixtures of ideal $\sanpone(\konea)$ and
$\onepone(\koneb)$ orbital angular momentum states, and their mixing
is given as
\eqn{eq::mixing}{%
\left(
  \begin{array}{c}
    \ket{\kkkl} \\
    \ket{\kkkh} \\
  \end{array}
\right) =\tetam\left(
  \begin{array}{c}
    \ket{\kkka} \\
    \ket{\kkkb} \\
  \end{array}\right), }
  where
\eqn{eq::mixingmatrix}{\tetam =\left(
                                 \begin{array}{cc}
                                   \sin\thetaK & \cos\thetaK \\
                                   \cos\thetaK & -\sin\thetaK \\
                                 \end{array}
                               \right)
} is the mixing matrix, and $\thetaK$ is the mixing
angle\cite{cenk}. The magnitude of the mixing angle is estimated to
be $34\degr\leq|\thetaK|\leq 58\degr$
\cite{cenk,Suzuki1,Burakovsky1997}. More recently, the sign of
$\thetaK$, and a new window for the value of $\thetaK$ is estimated
from the results of $B\rightarrow \kkkl\g$ and
$\tau\rightarrow\kkkl\nu_\tau$ data as\cite{hatanakayang}
\beq\label{eq::thetalimits} \thetaK = -(34\pm 13)\degr.\eeq In this
study, the results of \cite{hatanakayang} is used. Recently, the proporties of low lying meson poles are analyzed in \cite{MesRes}, indicating that there might be two meson poles corresponding to pseudo scalar $K_1(1270)$ states. In this work we assume the quark model picture, where there is only one pole, and no space for a second pole.

In this work, we calculate the transition form factors of $\toklh$
decays using three-point QCD sum rules approach. We also estimate
the branching fractions for these decays, with the final leptons
being $e^+e^-$, $\mu^+\mu^-$ and $\tau^+\tau^-$, and compare our
results with the ones in the literature.

This paper is organized as follows. In section $2$, the effective
Hamiltonian for $\toklh$ decays are defined. The sum rules for these
decays are derived in section $3$. The explicit expressions for the
form factors are also presented in section $3$. In section $4$, the
numerical analysis of the results, the estimated branching
fractions, and also the final discussions and remarks on the results
are given.

\section{Defining $B\rightarrow K_1(1270,1400)\ell^+ \ell^-$ transitions }

\begin{figure}[h]
\begin{center}
\includegraphics[width=8cm]{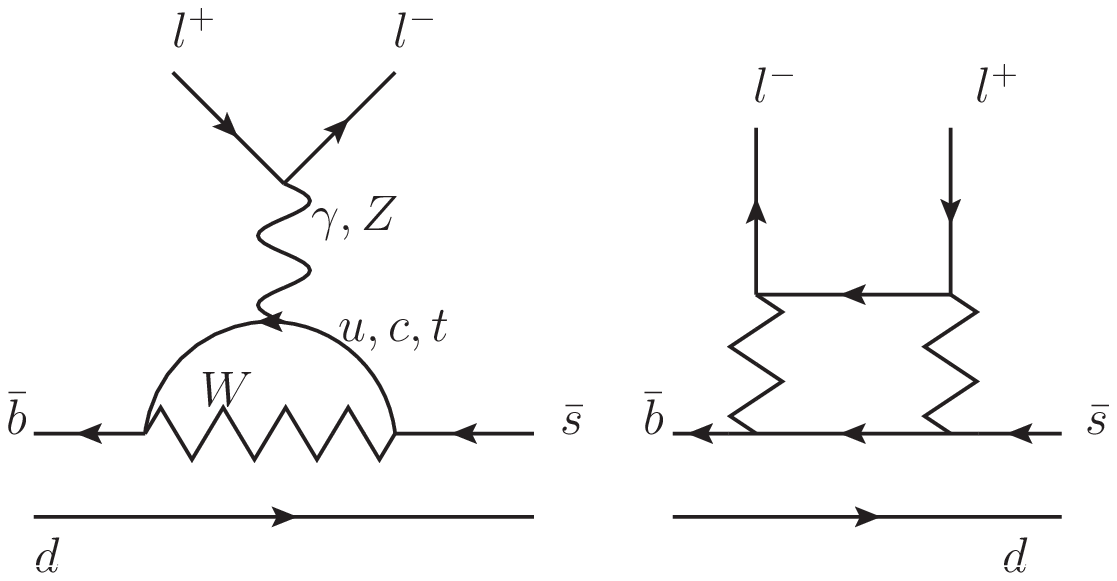}
 \vskip2mm \caption{\label{fig::diagrams}The loop penguin and box diagrams contributing to semileptonic $B$ to $K_1$ transitions.}
\end{center}
\end{figure}

In SM the $B\rightarrow K_1 \lp \lm$ transitions occur via
$b\rightarrow s \lp \lm$ loop transition, due to penguin and box
diagrams shown in Fig. \ref{fig::diagrams}. The effective
Hamiltonian for $b\rightarrow s \lp \lm$ transition is written
as\cite{gaur1}

\begin{eqnarray}\label{eq::ham}
\nn {\cal H}=\frac{G_F \alpha }{2 \sqrt{2} \pi} V_{tb}V^{*}_{ts}
&\times &\Bigg\{
C_9^{eff}\bar{s}\gamma_\mu(1-\gamma_5)b\bar{l}\gamma_\mu
l\\
\nn & +& C_{10} \bar{s}\gamma_\mu(1-\gamma_5)b\bar{l}\gamma_\mu
\gamma_5
l\\
&-&2 C_7^{eff} \frac{m^b}{q^2}
\bar{s}\sigma_{\mu\nu}q^\nu(1+\gamma_5)b\bar{l}\gamma_\mu l\Bigg\},
\end{eqnarray}
where $C_7^{eff}$, $C_9^{eff}$ and $C_{10}$ are the Wilson
coefficients, $G_F$ is the Fermi constant, $\alpha$ is the fine
structure constant at the $Z$ scale, $V_{ij}$ are the elements of
the CKM matrix and $q=p-p'$ is the momentum transferred to leptons.
By sandwiching the effective Hamiltonian in Eq. \ref{eq::ham}
between initial and final meson states, the transition amplitude for
$B\rightarrow K_1 \lp \lm$ decays is obtained as

\begin{eqnarray}\label{eq::amp}
\nn {\cal M}=\frac{G_F \alpha }{2 \sqrt{2} \pi} V_{tb}V^{*}_{ts}
&\times &\Bigg\{ C_9^{eff}\langle K_1
(p',\epsilon)|\bar{s}\gamma_\mu(1-\gamma_5)b|B(p)\rangle
\bar{l}\gamma_\mu
l\\
\nn & +& C_{10} \langle K_1
(p',\epsilon)|\bar{s}\gamma_\mu(1-\gamma_5)b|B(p)\rangle\bar{l}\gamma_\mu
\gamma_5
l\\
&-&2 C_7^{eff} \frac{m^b}{q^2} \langle K_1
(p',\epsilon)|\bar{s}\sigma_{\mu\nu}q^\nu(1+\gamma_5)b|B(p)\rangle\bar{l}\gamma_\mu
l\Bigg\},
\end{eqnarray}
where $p(p')$ is the momentum of the $B(K_1)$ meson, and $\epsilon$
is the polarization vector of the axial vector $K_1$ meson. In order
to calculate the amplitude, the matrix elements in Eq. \ref{eq::amp}
should be found. These matrix elements are parameterized in terms of
the form factors as

\begin{eqnarray}\label{eq::vectorforms}
\nn \langle K_1
(p',\epsilon)|\bar{s}\gamma_\mu(1-\gamma_5)b|B(p)\rangle & = &
\frac{2 i
A(q^2)}{M+m}\varepsilon_{\mu\nu\alpha\beta}\epsilon^{*\nu}p^\alpha p^{\prime\beta}- V_1{q^2}(M+m)\epsilon^{*}_{\mu}\\
&+&\frac{V_2(q^2)}{M+m}(\epsilon^{*}.p)P_\mu +
\frac{V_3(q^2)}{M+m}(\epsilon^{*}.p)q_\mu~~,~~
\end{eqnarray}
\begin{eqnarray}\label{eq::tensorforms}
\nn \langle K_1
(p',\epsilon)|\bar{s}\sigma_{\mu\nu}q^\nu(1+\gamma_5)b|B(p)\rangle&=&
2
T_1(q^2)\varepsilon_{\mu\nu\alpha\beta}\epsilon^{*\nu}p^\alpha p^{\prime\beta}\\
\nn &-&i T_2(q^2) [(M^2-m^2)\epsilon^{*}_{\mu} -
(\epsilon^{*}.p)P_\mu ] \\
&-&i T_3(\qsq)(\epsilon^{*}.p)\Bigg[q_\mu - \frac{\qsq
P_\mu}{M^2-m^2}\Bigg]~~,~~
\end{eqnarray}
where $P=p+p'$, $M\equiv M_B$, the mass of the $B$ meson and
$m\equiv m_{K_1}$ is the mass of the $K_1$ meson. The Dirac identity
\begin{equation}\label{eq::dirac}
\sigma_{\mu\nu}\gamma_5 =
\frac{-i}{2}\varepsilon_{\mu\nu\alpha\beta}\sigma_{\alpha\beta}
\end{equation}
with the convention $\gamma_5=i\gamma_0 \gamma_1 \gamma_2 \gamma_3$
and $\varepsilon_{0123}=-1$ requires that $T_1(0)=T_2(0)$. The
relation of the chosen form factors with the ones in the literature
\cite{hatanakayang,gaur1,cenk} are presented in table
\ref{tb::formliterature}.

\begin{table}[htb]\vskip1cm\caption{\label{tb::formliterature}The relation of form
factors used in this work, and used in
literature\cite{hatanakayang,gaur1,cenk}.}
\begin{center}
\begin{tabular}{cccc}
\hline\hline this work & \cite{hatanakayang} & \cite{gaur1} &\cite{cenk} \\
\hline $~~~~A$ & $A$ & $g(M+m)$& ~~~~~~~~\\
\hline $~~~~V_1$ & $V_1$ & $f/(M+m)$& \\
\hline $~~~~V_2$ & $V_2$ & $-a_+ (M+m)$ &\\
\hline $~~~~V_3$ & ~~~~$\frac{-2m(M+m)}{\qsq}(V_3 - V_0)$&$-a_- (M+m)$ ~~~~&\\
\hline $~~~~T_1$ & $T_1$&$-g_+$&$-Y_1 /2$\\
\hline $~~~~T_2$ & $T_2$&$-g_+ - g_- \frac{\qsq}{M+m}$&$Y_2$\\
\hline $~~~~T_3$ & $T_3$&$g_- + h(M+m)$&$Y_2$\\
\hline\hline
\end{tabular}

\end{center}
\vskip1cm
\end{table}

In this work the branching fractions of $\toklh$ transitions are
also estimated. The partial decay width of the $B$ meson is found by
squaring the amplitude in Eq. \ref{eq::amp}, and by multiplying the
phase space factors as
\begin{equation}\label{eq::partialwidth}
\frac{d\Gamma}{d\hat{q}} = \frac{G_{F}^2 \alpha^2 M}{2^{14} \pi^5} |
V_{tb}V_{ts}^{*} |^2 \lambda^{1/2}(1,\hat{r},\hat{q}) v
\Delta(\hat{q})~,
\end{equation}
where $\hat{q}=\qsq/M^2$, $ \lambda(a,b,c)=a^2+b^2+c^2-2(ab+bc+ca)$
and
\begin{eqnarray}\label{eq::pw2}
\Delta(\hat{q}) &=& \frac{2}{3 \hat{r} \hat{q}} M^2
 Re\Big[ - 12 M^2 \hat{m}_l \hat{q} \lambda(1,\hat{r},\hat{q}) \Big\{
({\cal E}_3-{\cal D}_2-{\cal D}_3) {\cal E}_1^\ast\nonumber \\ & -&
({\cal E}_2+{\cal E}_3-{\cal D}_3) {\cal D}_1^\ast \Big\} +12 M^4
\hat{m}_l \hat{q} (1-\hat{r}) \lambda(1,\hat{r},\hat{q})
({\cal E}_2-{\cal D}_2) ({\cal E}_3^\ast-{\cal D}_3^\ast) \nonumber \\
&+& 48 \hat{m}_l \hat{r} \hat{q} \Big\{ 3 {\cal E}_1 {\cal D}_1^\ast
+ 2 M^4 \lambda(1,\hat{r},\hat{q}) {\cal E}_0 {\cal D}_0^\ast \Big\}
- 16 M^4 \hat{r}\hat{q} (\hat{m}_l-\hat{q})
\lambda(1,\hat{r},\hat{q})
\Big\{ | {\cal E}_0|^2 + | {\cal D}_0|^2 \Big\} \nonumber \\
&-& 6 M^4 \hat{m}_l \hat{q} \lambda(1,\hat{r},\hat{q}) \Big\{ 2
(2+2\hat{r}-\hat{q}) {\cal E}_2 {\cal D}_2^\ast -
\hat{q} | ({\cal E}_3-{\cal D}_3)|^2 \Big\} \nonumber \\
&- &4 M^2 \lambda(1,\hat{r},\hat{q}) \Big\{ \hat{m}_l (2 - 2 \hat{r}
+ \hat{q} ) + \hat{q} (1 - \hat{r} - \hat{q} ) \Big\}
({\cal E}_1 {\cal E}_2^\ast + {\cal D}_1 {\cal D}_2^\ast) \nonumber\\
&+& \hat{q} \Big\{ 6 \hat{r} \hat{q} (3+v^2) +
\lambda(1,\hat{r},\hat{q}) (3-v^2) \Big\} \Big\{ | {\cal E}_1|^2 + |
{\cal D}_1|^2 \Big\}\nonumber\\&-& 2 M^4 \lambda(1,\hat{r},\hat{q})
\Big\{ \hat{m}_l [\lambda(1,\hat{r},\hat{q}) - 3 (1-\hat{r})^2] -
\lambda(1,\hat{r},\hat{q}) \hat{q} \Big\} \Big\{ | {\cal E}_2|^2 + |
{\cal D}_2|^2 \Big\}\Big]~,
\end{eqnarray}
and  $\hat{r}=m^2/M^2$, $\hat{m}_l=m_l^2/M^2$ and
$v=\sqrt{1-4\hat{m}_l/\hat{q}}$ is the final lepton velocity. The
following definitions are also used.
\begin{eqnarray} \label{eq::pwdefs}
{\cal D}_0 &=& (C_9^{eff}+C_{10}) \frac{A(q^2)}{M+m} +
 (2m_bC_7^{eff}) \frac{\ftb(q^2)}{q^2} ~, \nonumber\\
{\cal D}_1 &=& (C_9^{eff}+C_{10}) (M+m) \fvb(q^2) + (2m_bC_7^{eff})
(M^2-m^2)
\frac{T_{2}(q^2)}{q^2} ~, \nonumber\\
{\cal D}_2 &=& \frac{C_9^{eff}+C_{10}}{M+m} \fvi(q^2)
+(2m_bC_7^{eff}) \frac{1}{q^2}  \left[ T_{2}(q^2) +
\frac{q^2}{M^2-m^2} T_{3}(q^2) \right]~,
\nonumber\\
{\cal D}_3 &=& (C_9^{eff}+C_{10})\frac{\fvu(q^2)}{M+m} -
 (2m_bC_7^{eff}) \frac{T_{3}(q^2)}{q^2} ~, \nonumber\\
{\cal E}_0 &=& (C_9^{eff}-C_{10}) \frac{A(q^2)}{M+m} +
 (2m_{b}C_7^{eff}) \frac{\ftu(q^2)}{q^2} ~, \nonumber\\
{\cal E}_1 &=& (C_9^{eff}-C_{10}) (M+m) \fvb(q^2) +
(2m_{b}C_7^{eff}) (M^2-m^2)
\frac{T_{2}(q^2)}{q^2} ~, \nonumber\\
{\cal E}_2 &=& \frac{C_9^{eff}-C_{10}}{M+m} \fvi(q^2)
+(2m_{b}C_7^{eff}) \frac{1}{q^2}  \left[ T_{2}(q^2) +
\frac{q^2}{M^2-m^2} T_3(q^2) \right]~,
\nonumber\\
{\cal E}_3 &=& (C_9^{eff}-C_{10})\frac{\fvu(q^2)}{M+m} -
 (2m_{b}C_7^{eff}) \frac{T_3(q^2)}{q^2} . \nonumber\\
 \end{eqnarray}


%
\section{Sum rules for $\toklh$ transitions}

In this section the sum rules for the form factors of $\toklh$
transitions are found. In QCD sum rules approach, to obtain the
matrix elements in Eqs. \ref{eq::vectorforms} and
\ref{eq::tensorforms}, one can start from the three-point
correlation functions

\beqa\label{eq::correlatorA}
 \nn \Pi^{A,a}_{\mu\nu}(p^2,p^{\prime2})& =& i^2 \int dx^4 dy^4
 e^{-ipx} e^{ip^{\prime}y}\bra{0}T[J^{A}_\nu(y) J^{a}_\mu(0)
 J_{B}^\dagger(x)]\ket{0},\\
 \Pi^{T,a}_{\mu\nu\rho}(p^2,p^{\prime2})& =& i^2 \int dx^4 dy^4
 e^{-ipx} e^{ip^{\prime}y}\bra{0}T[J^{T}_{\nu\rho}(y) J^{a}_\mu(0)
 J_{B}^\dagger(x)]\ket{0},
\eeqa
where $J^{A}_\nu=\bar{s}\g_\nu\g_5 d$ and
$J^{T}_{\nu\rho}=\bar{s}\sigma_{\nu\rho}\g_5 d$ are axial vector and
tensor interpolating currents creating $K_1$ states, $J_{B}=\bb \g_5
d$ is the interpolating current of $B$ mesons, and
$J^{a}_\mu=J_\mu^{V-A,T+PT}$ are the vector and tensor parts of the
transition currents with $J_\mu^{V-A}=\bb\g_\mu(1-\g_5)s$ and
$J^{T+PT}=\bb\sigma_{\mu\vrr}q^\vrr (1+\g_5)s$.

The correlators are calculated in the following way. First, they are
saturated with two complete sets of intermediate states with same
quantum numbers of the initial and final state currents. These
calculations in terms of the matrix elements of $\kkkl$ and $\kkkh$
states form the phenomenological part of the QCD sum rules. The
phenomenological parts of the correlators (Eq.
\ref{eq::correlatorA}) can be written as

 \beqa\label{eq::phen1}
 \nn \corone &=&-\frac{\bra{0}J^{A}_\nu \ket{\kkkl \ppeps}\bra{\kkkl \ppeps} J^{a}_\mu\ket{B(p)}\bra{B(p)}J_B \ket{0}}{R_1 R}\\
 \nn & -&\frac{\bra{0}J^{A}_\nu \ket{\kkkh \ppeps}\bra{\kkkh \ppeps} J^{a}_\mu\ket{B(p)}\bra{B(p)}J_B \ket{0}}{R_2 R}\\
  \nn &+&\mbox{higher resonances and continuum states},\\
 \nn \cortwo &=&-\frac{\bra{0}J^{T}_{\nu\rho} \ket{\kkkl \ppeps}\bra{\kkkl \ppeps} J^{a}_\mu\ket{B(p)}\bra{B(p)}J_B \ket{0}}{R_1 R}\\
 \nn &-&\frac{\bra{0}J^{T}_{\nu\rho} \ket{\kkkh \ppeps}\bra{\kkkh \ppeps} J^{a}_\mu\ket{B(p)}\bra{B(p)}J_B \ket{0}}{R_2 R}\\
&+&\mbox{higher resonances and continuum states},
\eeqa
where $R=p^2-M^2$, $R_1=\ppsq-m_{\kkkl}^2$ and
$R_B=\ppsq-m_{\kkkh}^2$. The matrix elements for the $B$ meson is
defined as

\beq\label{e::formBmes} \bra{B(p)}J_B \ket{0}=-i\frac{F_B M^2}{m_b +
m_d}. \eeq

In QCD sum rules, each correlator function has its own continuum.
Due to this fact, obtaining the matrix elements $\bra{\kkkl \ppeps}
J^{a}_\mu\ket{B(p)}$ and $\bra{\kkkh \ppeps} J^{a}_\mu\ket{B(p)}$
from two correlator reduces the reliability of the sum rules. An
alternative way to obtain the transition matrix elements is to
express $\kkkh$ and $\kkkh$ states in terms of $\kkka$ and $\kkkb$
which are G-parity eigenstates as defined in Eq.
\ref{eq::mixing}\cite{yang87,hatanakayang}.In this work, G-parity is used as a generalization of C parity defined for $q\bar{q}$ mesons to $q \bar{q'}$ multiplets as done in\cite{yang87}.

The matrix elements  $\Mtokl$ and $\Mtokh$ in Eq. \ref{eq::amp} can
be written in terms of matrix elements $\Mtoka$ and $\Mtokh$ states
as\cite{yangana}

\beq\label{eq::matrixelementmix} \left(
  \begin{array}{c}
    \bra{\kkkl}J_\mu\ket{B} \\
    \bra{\kkkh}J_\mu\ket{B} \\
  \end{array}
\right)=\tetam \left(
                 \begin{array}{c}
                   \bra{\kkka}J_\mu\ket{B} \\
                    \bra{\kkkb}J_\mu\ket{B} \\
                 \end{array}
               \right)
\eeq
where $J_\mu$ are any of the transition currents. Due to this
relation, the form factors parameterizing $\Mtoklh$ matrix elements
can be expressed in terms of the form factors parameterizing
$\Mtokab$ matrix elements as follows

\beq\label{eq::formmixing} \left(
  \begin{array}{c}
    \xi \ffl \\
    \xi' \ffh \\
  \end{array}
\right)=\tetam\left(
  \begin{array}{c}
    \varsigma \ffa \\
    \varsigma' \ffb \\
  \end{array}
\right)
\eeq where $f_i$ is defined as the form factors
$\{A,V_1,V_2,V_3,T_1,T_2,T_3\}$ respectively for $i=1,2,...,7$, and
$\ffl$, $\ffh$, $\ffa$ and $\ffb$ denotes the form factors
parameterizing $\Mtokl$, $\Mtokh$, $\Mtoka$ and $\Mtokb$ matrix
elements respectively. The values for factors $\xi$, $\xi'$,
$\varsigma$ and $\varsigma'$ are given in table
\ref{tb::factorsforms}, where $m_1\equiv m_{\kkkl}$, $m_2\equiv
m_{\kkkh}$, $m_A\equiv m_{\kkka}$ and $m_B\equiv m_{\kkkb}$. The
masses of $\kkka$ and $\kkkb$ states are defined
as\cite{hatanakayang}

\beqa\label{ch3::eqMab} \nn m_{K_{1A}}^2 &=& m_{K_1(1400)}^2
\cos^2\theta_K + m_{K_1(1270)}^2
 \sin^2\theta_K \,\\
 m_{K_{1B}}^2 &=&
 m_{K_1(1400)}^2 \sin^2\theta_K + m_{K_1(1270)}^2 \cos^2\theta_K .\eeqa

\begin{table}[htb]\vskip1cm\caption{\label{tb::factorsforms}The
values for factors $\xi$, $\xi'$, $\varsigma$ and $\varsigma'$ for
the form factors.}
\begin{center}

\begin{tabular}{ccccc}
\hline\hline ~~~~$f_i$~~~~ &~~~~$\xi$~~~~&~~~~$\xi'$~~~~&~~~~$\varsigma$~~~~&~~~~$\varsigma'$~~~~\\
\hline $A,V_2,V_3$&$1/(M+m_1)$&$1/(M+m_2)$&$1/(M+m_A)$&$1/(M+m_B)$\\
\hline $V_1$&$(M+m_1)$&$(M+m_2)$&$(M+m_A)$&$(M+m_B)$\\
\hline $T_1,T_3$&$1$&$1$&$1$&$1$\\
\hline $T_2$ &$(M^2-m^2_1)$&$(M^2-m^2_2)$&$(M^2-m^2_A)$&$(M^2-m^2_B)$\\
\hline\hline
\end{tabular}
\end{center}\vskip1cm
\end{table}

Inserting Eqs. \ref{eq::mixing} and  \ref{eq::matrixelementmix} in
Eq. \ref{eq::phen1}, and applying double Borel transformations, the
phenomenological parts of the correlators are found in terms of
G-parity eigen states as

 \beqa\label{eq::phen1XX}
 \nn \coroneB &=&-e^{\frac{-M^2}{M_1^2}}e^{\frac{-m_1^2}{M_2^2}}\Bigg\{\bra{0}J^{A}_\nu \Bigg[s^2 \ket{\kkka \ppeps}\bra{\kkka \ppeps}+c^2\ket{\kkkb \ppeps}\bra{\kkkb \ppeps}\\
 \nn &&+sc \Bigg(\ket{\kkka \ppeps}\bra{\kkkb \ppeps}+\ket{\kkkb \ppeps}\bra{\kkka \ppeps}\Bigg)\Bigg] J^{a}_\mu\ket{B(p)}\bra{B(p)}J_B \ket{0}\Bigg\}\\
 \nn & &-e^{\frac{-M^2}{M_1^2}}e^{\frac{-m_2^2}{M_2^2}}\Bigg\{\bra{0}J^{A}_\nu \Bigg[c^2 \ket{\kkka \ppeps}\bra{\kkka \ppeps}+s^2\ket{\kkkb \ppeps}\bra{\kkkb \ppeps}\\
 \nn &&-sc \Bigg(\ket{\kkka \ppeps}\bra{\kkkb \ppeps}+\ket{\kkkb \ppeps}\bra{\kkka \ppeps}\Bigg)\Bigg] J^{a}_\mu\ket{B(p)}\bra{B(p)}J_B \ket{0}\Bigg\}\\
 \nn \cortwoB &=&-e^{\frac{-M^2}{M_1^2}}e^{\frac{-m_1^2}{M_2^2}}\Bigg\{\bra{0}J^{T}_{\nu\rho} \Bigg[s^2 \ket{\kkka \ppeps}\bra{\kkka \ppeps}+c^2\ket{\kkkb \ppeps}\bra{\kkkb \ppeps}\\
 \nn &&+sc \Bigg(\ket{\kkka \ppeps}\bra{\kkkb \ppeps}+\ket{\kkkb \ppeps}\bra{\kkka \ppeps}\Bigg)\Bigg] J^{a}_\mu\ket{B(p)}\bra{B(p)}J_B \ket{0}\Bigg\}\\
 \nn & &-e^{\frac{-M^2}{M_1^2}}e^{\frac{-m_2^2}{M_2^2}}\Bigg\{\bra{0}J^{T}_{\nu\rho} \Bigg[c^2 \ket{\kkka \ppeps}\bra{\kkka \ppeps}+s^2\ket{\kkkb \ppeps}\bra{\kkkb \ppeps}\\
\nn &&-sc \Bigg(\ket{\kkka \ppeps}\bra{\kkkb \ppeps}+\ket{\kkkb
\ppeps}\bra{\kkka \ppeps}\Bigg)\Bigg]
J^{a}_\mu\ket{B(p)}\bra{B(p)}J_B
 \ket{0}\Bigg\},\\
 &&
\eeqa
where $s\equiv \sin\thetaK$ and $c\equiv\cos\thetaK$. $M_1^2$ and
$M_2^2$ appearing in Eq. \ref{eq::phen1XX} are Borel mass parameters
and $\Bpi$ denotes the Borel transformation of $\Pi$. The double
Borel transformation with respect to the variables $p^2$ and $p'^2$
($p^2\rightarrow M_{1}^2,p'^2\rightarrow M_{2}^2$) is given as
\begin{equation}\label{eq::boreltrans}
\hat{{\cal
B}}\Bigg[\frac{1}{(p^2-m^2_1)^m}\frac{1}{(p'^2-m^2_2)^n}\Bigg]\rightarrow(-1)^{m+n}\frac{1}{\Gamma(m)}\frac{1}{\Gamma
(n)}e^{-m_{1}^2/M_{1}^{2}}e^{-m_{2}^2/M_{2}^{2}}\frac{1}{(M_{1}^{2})^{m-1}(M_{2}^{2})^{n-1}}.
\end{equation}

The matrix elements of $\kkkab$  states are defined in terms of both
G parity conserving and violating decay constants discussed in
\cite{yangana}. The G parity conserving decay constants are given as
\beqa\label{eq::gparitycons}
 \nn \bra{\kkka (p',\epsilon)}\bar{s}\g_\mu\g_5 d\ket{0}& = & i
 f_{\kkka}
 m_A \epsilon_\mu^*,\\
\bra{\kkkb (p',\epsilon)}\bar{s}\sigma_{\mu\nu} \g_5 d\ket{0}& = &
f_{\kkkb}^{\perp}[\epsilon^*_\mu p'_\nu-\epsilon^*_\nu p'_\mu],
\eeqa
and the G parity violating decay constants are given as
\beqa\label{eq::gparityviol}
 \nn \bra{\kkka (p',\epsilon)}\bar{s}\sigma_{\mu\nu}\g_5 d\ket{0}& = & i
 f_{\kkka}
 a_0^{\perp \kkka} [\epsilon^*_\mu p'_\nu-\epsilon^*_\nu p'_\mu],\\
\bra{\kkkb (p',\epsilon)}\bar{s}\g_\mu\g_5 d\ket{0}& = & i
f_{\kkkb}^{\perp} m_B (1GeV)a_0^{\parallel\kkkb}\epsilon_\mu^*,
\eeqa
where $f_{\kkka}(\equiv f_A)$ and $f_{\kkkb}^{\perp}(\equiv f_B )$
are the decay constants of $\kkka$ and $\kkkb$ mesons, and
$a_0^{\perp \kkka}$ and $a_0^{\parallel\kkkb}$ are the zeroth
Gegenbauer moments. Since the Gegenbauer moments are zero in $SU(3)$
limit\cite{yang87}, the G parity violating matrix elements are
expected to be small. In \cite{yangana}, their values are predicted
to be consistent with zero. In this work, they will be neglected.
After defining the matrix elements $\Mtokab$ and inserting in Eq.
\ref{eq::phen1XX} the following assumptions are made.

\beqa\label{eq::assAB} \nn &&e^{\frac{-m_1^2}{M_2^2}}s^2 \ket{\kkka
\ppeps}\bra{\kkka \ppeps}  + e^{\frac{-m_2^2}{M_2^2}}c^2 \ket{\kkka
\ppeps}\bra{\kkka \ppeps} \sim e^{\frac{-m_A^2}{M_2^2}} \ket{\kkka
\ppeps}\bra{\kkka \ppeps}\\
\nn &&e^{\frac{-m_1^2}{M_2^2}}c^2 \ket{\kkkb \ppeps}\bra{\kkkb
\ppeps} + e^{\frac{-m_2^2}{M_2^2}}s^2 \ket{\kkkb \ppeps}\bra{\kkkb
\ppeps} \sim e^{\frac{-m_B^2}{M_2^2}} \ket{\kkkb
\ppeps}\bra{\kkkb \ppeps}\\
\nn &&(e^{\frac{-m_1^2}{M_2^2}}- e^{\frac{-m_2^2}{M_2^2}})sc
\Bigg(\ket{\kkka \ppeps}\bra{\kkkb \ppeps}+\ket{\kkkb
\ppeps}\bra{\kkka \ppeps}\Bigg)\sim 0 .\\
&& \eeqa

The numerical values of the masses of $K_1$ states given in
numerical discussions satisfy $m_1<m_A<m_B<m_2$. And also the
minimum value of the Borel mass parameter $M_2^2$ guarantees
$e^{\frac{-m_1^2+m_2^2}{M_2^2}}>0.94$. Due to this considerations
the assumptions made in Eq. \ref{eq::assAB} effects the results of
the form factors by less than $3\%$. After employing the assumptions
defined in Eq. \ref{eq::assAB}, the phenomenological parts of the
correlators are written in terms of G-parity eigenstates as
 \beqa\label{eq::phen1AB}
 \nn \coroneB &=&-e^{\frac{-M^2}{M_1^2}}e^{\frac{-m_A^2}{M_2^2}}\bra{0}J^{A}_\nu \ket{\kkka \ppeps}\bra{\kkka \ppeps} J^{a}_\mu\ket{B(p)}\bra{B(p)}J_B
 \ket{0}\\
 \nn \cortwoB &=&-e^{\frac{-M^2}{M_1^2}}e^{\frac{-m_B^2}{M_2^2}}\bra{0}J^{T}_{\nu\rho} \ket{\kkkb \ppeps}\bra{\kkkb \ppeps} J^{a}_\mu\ket{B(p)}\bra{B(p)}J_B \ket{0}.\\
 &&
\eeqa
Using equations \ref{e::formBmes}, \ref{eq::gparitycons} and
\ref{eq::phen1AB} and summing over the polarizations of the $\kkkab$
mesons, the so called phenomenological parts of the correlation
functions are found and expressed in terms of selected structures as
\beqa\label{eq:structuresAphen}
 \nn \Bpi^{A(V-A)}_{\mu\nu} &=& \frac{F_B
M^2}{m_b + m_c}f_A m_A
e^{\frac{-M^2}{M_1^2}}e^{\frac{-m_A^2}{M_2^2}}\Bigg[g_{\mu\nu}\faa
(M+m_A)\\
\nn&&+\frac{1}{2}\fvia (M+m_A)(p_\mu p_\nu +p'_\mu
p_\nu)\\
 &&+ \frac{1}{2}\fvua(M+m_A)(p_\mu p_\nu -p'_\mu p_\nu)\\
 \nn&&+ i\frac{\fvba\epss{\mu}{\nu}{\rho}{\vrr}p^{\rho}\ppu{\vrr}}{(M+m_A)}\Bigg],\\
 \nn \Bpi^{A(T+PT)}_{\mu\nu} &=& \frac{F_B
M^2}{m_b + m_c}f_A m_A
e^{\frac{-M^2}{M_1^2}}e^{\frac{-m_A^2}{M_2^2}}\Bigg[
i\ftba\epss{\mu}{\nu}{\rho}{\vrr}p^{\rho}\ppu{\vrr}\\
\nn&&+\frac{\ftia g_{\mu\nu}}{M^2-m^2_A}+\ftua(p_\mu p_\nu +p'_\mu
p_\nu)/2\Bigg], \eeqa and \beqa\label{eq::structuresBphen}
 \nn \Bpi^{T(V-A)}_{\mu\nu\rho} &=& i \frac{F_B
M^2}{m_b + m_c}f_B e^{\frac{-M^2}{M_1^2}}e^{\frac{-m_B^2}{M_2^2}}
\Bigg[\fab (M+m_B)
g_{\mu\nu}p'_{\rho}\\
\nn && +\frac{1}{2}\fvib (M+m_B)(p_\mu p_\nu +p'_\mu
p_\nu)p_{\rho}\\
 \nn &&+\frac{1}{2} \fvub(M+m_B)(p_\mu p_\nu -p'_\mu p_\nu)p_{\rho}
 \\
 \nn &&+
i\frac{\fvbb\epss{\mu}{\nu}{\alpha}{\vrr}p^{\alpha}\ppu{\vrr}p_{\rho}}{(M+m_B)}
\Bigg],\\
\nn \Bpi^{T(T+PT)}_{\mu\nu\rho} &=& \frac{F_B M^2}{m_b + m_c}f_B
e^{\frac{-M^2}{M_1^2}}e^{\frac{-m_B^2}{M_2^2}} \Bigg[i\frac{1}{2}
\ftbb\epss{\mu}{\nu}{\alpha}{\vrr}p^{\alpha}\ppu{\vrr}p_{\rho}\\
\nn && +\frac{\ftib
g_{\mu\nu}p_{\rho}}{(M^2-m^2_B)}\\
&&+\frac{1}{2}\ftub(p_\mu p_\nu +p'_\mu p_\nu)p_{\rho}\Bigg].
\eeqa

In QCD sum rules, the correlation functions are also calculated
theoretically using the operator product expansion (OPE) in the
space-like region where $p'^2 \ll (m_{s}+m_{d})^2 $ and $p^2 \ll
(m_{b}+m_d)^2$ in the so called deep Euclidean region. The
contributions to the correlation functions in the QCD side of sum
rules come from bare-loop (perturbative) diagrams and also quark
condensates (nonperturbative). The correlators can be written as

\beqa\label{eq:structuresA}
 \nn \Bpi^{A(V-A)}_{\mu\nu} &=& \Bpi_{\faa}
g_{\mu\nu}\\
\nn && +\frac{\Bpi_{\fvia}(p_\mu p_\nu +p'_\mu p_\nu)}{2} +
\frac{\Bpi_{\fvua}(p_\mu p_\nu -p'_\mu p_\nu)}{2}\\
&& + i\Bpi_{\fvba}\epss{\mu}{\nu}{\rho}{\vrr}p^{\rho}\ppu{\vrr},\\
 \nn \Bpi^{A(T+PT)}_{\mu\nu} &=&
\Bpi_{\ftba}\epss{\mu}{\nu}{\rho}{\vrr}p^{\rho}\ppu{\vrr}+\Bpi_{\ftia}g_{\mu\nu}\\
\nn &&+\Bpi_{\ftua}\frac{(p_\mu p_\nu +p'_\mu p_\nu)}{2},
\eeqa and \beqa\label{eq::structuresB}
 \nn \Bpi^{T(V-A)}_{\mu\nu\rho} &=&  i\Bpi_{\fvib}\frac{(p_\mu
p_\nu +p'_\mu p_\nu )p_{\rho}}{2}\\
\nn && +
\frac{\Bpi_{\fvub}(p_\mu p_\nu -p'_\mu p_\nu)p_{\rho}}{2}\\
\nn && +\Bpi_{\fab} g_{\mu\nu}p'_{\rho}+ i\Bpi_{\fvbb}\epss{\mu}{\nu}{\alpha}{\vrr}p^{\alpha}\ppu{\vrr}p_{\rho},\\
\nn \Bpi^{T(T+PT)}_{\mu\nu\rho}& =& i\frac{
\Bpi_{\ftbb}\epss{\mu}{\nu}{\alpha}{\vrr}p^{\alpha}\ppu{\vrr}p_{\rho}}{2}\\
\nn  &&+\Bpi_{\ftib}g_{\mu\nu}p_{\rho}\\
&&+\frac{\Bpi_{\ftub}(p_\mu p_\nu +p'_\mu p_\nu)p_{\rho}}{2}.
\eeqa Each of $\Bpi_{f_{i(A,B)}}$ are expressed in terms of
perturbative and nonperturbative contributions as

\beq\label{eq::pert1}
\Bpi_{f_{i(A,B)}}=\Bpi_{f_{i(A,B)}}^{pert}+\Bpi_{f_{i(A,B)}}^{nonpert}.
\eeq
The perturbative parts of the correlators are written in terms of
double dispersion relation for the coefficients of the selected
Lorentz structures, as

\begin{equation}\label{eq::doubledisp}
\Bpi_{f_i}^{per}=\int ds\int
ds'\rho_{f_i}(s,s',q^2)e^{\frac{-s}{M_1^2}}e^{\frac{-s'}{M_2^2}},
\end{equation}

where $\rho_{f_i}(s,s',q^2)$ are the spectral densities defined as

\beq \label{eq::spectraldef} \rho_{f_i}(s,s';q^2) = \frac{ Im_{s}
Im_{s'} \Pi_{f_i}^{OPE}(s,s';q^2)}{\pi^2}.\eeq

The spectral densities in Eq. \ref{eq::doubledisp} are calculated by
using the usual Feynman integral for the loop diagrams, with the
help of Cutkovsky rules, i.e., by inserting delta functions instead
of the quark propagators ($\frac{1}{p^2-m^2}\rightarrow 2 i \pi
\delta(p^2-m^2)$), implying that all quarks are real. The physical
region in $s,s'$ plane is described by the following inequality

\begin{equation}\label{eq::inequality}
-1\leq f(s,s')=\frac{2ss'+ (m_b^2-s-m_d^2)(s+s'-q^2)+2
s(m_{b}^2-m_{d}^2)}
{\lambda^{1/2}(m_{b}^2,s,m_{d}^2)\lambda^{1/2}(s,s',q^2)}\leq+1.
\end{equation}
The calculations lead to the following results for the spectral
densities. For the $\Mtoka$ matrix elements, the spectral densities
are calculated as

 \beqa\label{ee::spactralA}
 \rho_{A_{A}}&=& 2 (M+m)
 I_0 \{m_d +(-m_b+ m_d)A_1 + (m_d + m_s)B_1\}
 ,\\
  \rho_{V_{1A}}&=& \frac{2}{M+m}
 I_0 \{m_d [(m_d-m_b)(m_d+m_s)-q^2 +s +s']\\
 \nn &&+[2 m_s s + m_b (q^2 - s - s')+ m_d(-q^2 +3s +s')]A_1 + 4(m_b-m_d)A_2 \\
 \nn && + [(m_d + m_s)(s-q^2)+(m_s + 3 m_d - 2 m_b)]B_1\} ,\\
  \rho_{V_{2A}}&=& 2 (M+m)
 I_0 \{m_d - (m_b-3 m_d)A_1 + (m_d + m_s)B_1\\
 \nn && - 2 (m_b - m_d)(B_2 + D_2)\} ,\\
  \rho_{V_{3A}}&=& -2 (M+m)
 I_0 \{m_d - (m_b+ m_d)A_1 + (m_d + m_s)B_1\\
 \nn && + 2 (m_b - m_d)(B_2 - D_2)\} ,\\
  \rho_{T_{1A}}&=& -4
 e^{\frac{-s}{M_1^2}}e^{\frac{-s'}{M_2^2}}I_0 \{m_d (m_s - m_b) +6 A_2\\
 \nn &&+[s+(m_d-m_s)(m_s+m_d)]A_1 +[s'+(m_d-m_s)(m_s+m_d)]B_1\\
 \nn && + 2 s B_2 -(q^2 - s)(C_2 + D_2)+s' (C_2 + D_2 +2 F_2)\} \\
  \rho_{T_{2A}}&=& \frac{2}{M^2-m^2}
 I_0 \{-m_d [2 m_d (s-s') + m_s (q^2 +s - s' ) + m_b (q^2 -s + s') ]\\
 \nn && + [-m_d^2 (q^2 + s -s') - m_d m_s (q^2 +s-s') +m_b(m_d+m_s)(q^2+s-s')+s(q^2-s+s')]A_1\\
 \nn && + 2 (q^2 +s-s')A_2 +[(m_d-m_b)(m_d+m_s)(q^2-s+s')-(q^2 + s -s')s' ] B_1\} ,\\
  \rho_{T_{3A}}&=& 2
 I_0 \{m_d(2m_d-m_b+m_s)+[s+(m_d-m_b)(m_d+m_s)]A_1\\
 \nn &&- 2A_2 +[s'+(m_d-m_b)(m_d+m_s)]B_1+2 q^2 D_2\} .\eeqa
For the $\Mtokb$ matrix elements, the spectral densities are
calculated as %
\beqa\label{eq::spectralB}
 \rho_{A_{B}}&=& -8 (M+m)I_0(s,s',q^2)
 \{ B_1+D_2+F_2\},\\
 \rho_{V_{1B}}&=&
 \frac{4}{M+m}I_0(s,s',q^2)\{(m_b - m_d)m_d - s A_1 \\
 \nn && +[(m_b- m_d)(m_d + m_s)+q^2 -s - s']B_1 - 2 s D_2 + (q^2-s-s')F_2
 \},\\
  \rho_{V_{2B}}&=& -4 (M+m)I_0(s,s',q^2)
 \{ B_1+D_2+F_2\},\\
  \rho_{V_{3B}}&=& 4 (M+m)I_0(s,s',q^2)
 \{ B_1-D_2+F_2\},\\
   \rho_{T_{1B}}&=& 8I_0(s,s',q^2)
 \{ (m_b-m_s)(B_1+D_2+F_2)\},\\
  \rho_{T_{2B}}&=& \frac{-4}{M^2-m^2}
 I_0(s,s',q^2)\{[s'+(m_d-m_b)(m_d+m_s)-4 (m_b-m_d)A_2]\\
 \nn && +[s m_s +s' m_b +m_d(q^2 - 2s')]A_1 +s'(m_b-2 m_d+m_s)B_1\\
 \nn && +(m_d-m_b)(q^2 +s-s')B_2+(m_b-m_d)(q^2 -s+s')C_2 \},\\
   \rho_{T_{3B}}&=& -4
 I_0(s,s',q^2)\{ m_d - (m_b-2m_d)A_1 -2 (m_b-2m_d)B_1\\
 \nn && -(m_b-m_d)(B_2 +2D_2+F_2)\},
\eeqa where \beqa\label{eq::intCoefficients}
I_0(s,s',q^2)&=&\frac{1}{\lambda^{\frac{1}{2}}(s,s',q^2)},\\
A_1 &=& \frac{s'(q^2 + s-s'-2m_b^2)+m_d^2(q^2-s+s')+m_s^2(s+s'-q^2)}{q^4 -(s-s')^2-2 q^2(s+s')},\\
B_1 &=& \frac{s(q^2 - s+s'-2m_s^2)+m_d^2(q^2+s-s')+m_b^2(-s-s'+q^2)}{q^4-(s-s')^2-2q^2(s+s')},\\
A_2 &=& \frac{1}{2(q^4-(s-s')^2-2q^2(s+s'))}\{m_d^4 q^2 +m_b^4 s' +s(m_s^4 +q^2 s'-m_s^2(q^2-s+s'))\\
\nn && -m_b^2 [s'(q^2 +s-s') +m_d^2(q^2-s+s')+m_s^2(s+s'-q^2)]\\
\nn &&-m_d^2[m_s^2(q^2+s-s')+q^2(-q^2+s+s')]\},\\
B_2 &=&\frac{1}{(q^4-(s-s')^2-2q^2(s+s'))^2}\{  m_s^4 [q^4 +s^2 +4ss' +s^{\prime 2}-2 q^2 (s+s')]\\
\nn &&+ s^{\prime 2}[6 m_b^4 +q^4+4 s q^2 +s^2 -6m_b^2(q^2
+s-s')-2(q^2 +s)s'+s^{\prime 2}]\\
\nn &&+m_d^4 [(q^2-s)^2 + 4q^2 s' -2
ss'+s^{\prime 2}]\\
 \nn && -2m_s^2 s'[q^4 -2 s^2 +q^2(s-2s')+ s s'
+s^{prime 2}+3m_b^2(s+s'-q^2)]\\
\nn && -2 m_d^2 [m_s^2((s-q^2)^2+(s+q^2)s'-2s^{\prime 2})+s'(-2q^4
+(s-s')^2 \\
\nn &&~~~~~~~~+3m_b^2(q^2-s+s')+q^2(s+s'))] \},\\
C_2 &=&\frac{1}{(q^4-(s-s')^2-2q^2(s+s'))^2}\{ 3 m_b^4 (q^2- s - s' ) s' \\
\nn && -2 m_b^2[(m_d^2-m_s^2)(q^2-s)^2 + 2 m_s^2 s'(q^2 -2s)+s(m_d^2(q^2 + s)+(q^2-s)(q^2+2s))\\
\nn &&~~~~~~~~-s^{\prime 2}(2 m_d^2+m_s^2+2q^2-s)+s^{\prime 3}]\\
\nn && + m_d^4 [2 q^4 -(s-s')^2-q^2(s+s')]\\
\nn && - m_d^2[-q^6 +q^4(s+s')-(s-s')^2(s+s')+q^2(s^2-6s s' +
s^{\prime 2})\\
\nn && ~~~~~~~~+2 m_s^2(q^4 -2 s^2 +q^2(s-2s')+s s'+s^{\prime
2})]\\
\nn && -s[3 m_s^4(s+s'-q^2)+ 2 m_b^2((q^2-s)^2+(q^2-s)s'-2 s^{\prime
2})\\
\nn && ~~~~~~~~+s'(-2q^4+{s-s'}^2+q^2(s+s')) ]\},\\
D_2&=&C_2,\\
F_2 &=&\frac{1}{(q^4-(s-s')^2-2q^2(s+s'))^2}\{m_d^4[q^4 +q^2 s+s^2
-2s'(q^2-s)+s^{\prime 2}]\\
\nn && +s^2[6 m_s^4 +(q^2-s)^2+4 q^2 s' - 2ss' +s^{\prime 2}-6
m_s^2(q^2 -s +s')]\\
\nn && +m_b^4[q^4 +s^2 + 4 s s' + s^{\prime 2}-2 q^2(s+s')]\\
\nn && -2 m_d^2 s [-2 q^4 +(s-s')^2+3 m_s^2(q^2 + s-s')+q^2(s+s')]\\
\nn && -2m_b^2[m_d^2(q^4 -2s^2 + q^2 (s-2s')+ s s' + s^{\prime
2})\\
\nn && ~~~~~~~~ +s((q^2-s)^2+(q^2 +s)q^2 - 2 s^{\prime 2} + 3
m_s^2(s+s'-q^2))] \}. \eeqa

The nonperturbative contributions to the correlators are calculated
by taking the operators with dimensions $d=3(\qq)$ and
$d=5(m_0^2\qq)$ into account. For the $\Mtoka$ matrix elements
nonperturbative parts of the correlators are calculated as
\beqa\label{eq::npAA} \label{nonpertILK}\Pi_{A_{A}}&=& (M+m)\qq
\{\frac{1}{r r'}\}+m_0^2 (M+m)\qq \{\frac{1}{8 r
\rru{2}}-\frac{m_s^2}{2
r\rru{3}}\\
\nn &&-\frac{m_b^2}{2 r^3 r'}+\frac{1}{8 r^2
r'}+\frac{m_b^2-q^2}{r^2 \rru{2}}
\},\\
\Pi_{V_{1A}}&=&\frac{\qq}{M+m}\{ \frac{(m_b-m_s)^2-q^2}{2  rr'}\}\\
\nn &&+\frac{m_0^2 \qq}{M+m}\{\frac{(q^2 -(m_b-m_s)^2)m_s^2}{4 r \rru{3}} + \frac{(q^2 -(m_b-m_s)^2)m_b^2}{4 r^3 r'}\\
\nn && +\frac{m_b^2 + 7 m_b m_s -q^2}{8 r \rru{2}}+\frac{m_s^2 + 7 m_b m_s -q^2}{8 r^2 r}\\
\nn &&+\frac{((m_b-m_s)^2-q^2)(m_b^2+m_s^2-q^2)}{ r^2 \rru{2}}\},\\
\Pi_{V_{2A}}&=& (M+m)\qq\{ \frac{1}{2 r r'}\}\\
\nn &&- m_0^2 (M+m)\qq \{\frac{m_s}{4 r
\rru{3}}-\frac{1}{16 r\rru{2}}\\
\nn && +\frac{1}{16 r'r^2} +\frac{m_b}{4
r^3r'}\\
\nn &&+\frac{q^2 -m_s^2 -m_b^2}{16 r^2\rru{2}}
\},\\
\Pi_{V_{3A}}&=& -(M+m)\qq \{\frac{1}{2  r r'}\}\\
\nn &&m_0^2 (M+m)\qq \{\frac{m_s}{4 r \rru{3}} -\frac{1}{16
r\rru{2}} +\frac{3}{16
r'r^2} \\
\nn && +\frac{m_b}{4 r^3r'}+\frac{q^2 -m_s^2 -m_b^2}{16 r^2\rru{2}}
\},\\
\Pi_{T_{1A}}&=& -\qq\{\frac{(m_b)}{16 rr'}\} -m_0^2
\qq\{\frac{m_s^2(m_b-m_s)}{r \rru{3}}+\frac{m_b^2(m_b-m_s)}{r^3
r'}\\
\nn &&-\frac{(m_b+8 m_s)}{8 r \rru{2}}+\frac{(m_b+ m_s)}{8 r^2
r'}-\frac{(m_b-m_s)(m_b^2 -q^2)}{8 r^2 \rru{2}} \},\eeqa \beqa
\Pi_{T_{2A}}&=&  - \frac{ \qq}{M^2-m^2}\{\frac{(m_b)(8 m_b^2
-9m_s^2+8m_b(m_b-2m_s)-8q^2)}{
rr'}\}\\
\nn && - \frac{m_0^2 \qq}{M^2-m^2}\{\frac{m_b^2(m_b+m_s)(m_b^2-2 m_b
m_s -q^2)}{4 r^3
r'}\\
\nn &&-\frac{[2 m_b^3 + 7m_b^2 m_s - m_s(7
m_b^2-2m_s^2+7q^2)+2m_b(-m_s^2-q^2)]}{16 r \rru{2}}\\
\nn &&+\frac{[9 m_b^3+ 2m_s q^2 +m_b(-14 m_s^2+7q^2)]}{16 r^2
r'}\\
\nn &&-\frac{(m_b+m_s)(m_b^2 -q^2)(m_b^2 -2m_b
m_s-q^2)}{16 r^2 \rru{2}}\\
\nn &&+\frac{m_s^2(m_b+m_s)(m_b^2-2 m_b m_s -q^2)}{4 r
\rru{3}} \},\\
\Pi_{T_{3A}}&=&\qq \{\frac{m_b}{16  rr'}\}+m_0^2 \qq
\{\frac{m_s^2(m_b-m_s)}{4 r \rru{3}}+\frac{m_b^2(m_b-m_s)}{4 r^3
r'}\\
\nn &&+\frac{(8 m_b+ m_s)}{ 16 r^2 r'}-\frac{(m_b-m_s)(m_b^2
-q^2)}{16 r^2 \rru{2}}+-\frac{(m_b+8 m_s)}{8 r \rru{2}} \}.
 \eeqa
 For the $\Mtokb$ matrix elements the nonperturbative parts of the correlators are calculated as
\beqa\label{eq::npBB}
 \Pi_{A_{B}}&=& 0,\\
 \Pi_{V_{1B}}&=& \frac{ \qq}{M+m}\{\frac{m_b}{ r r'}\}-\frac{m_0^2 \qq}{M+m}\{\frac{m_s^2 m_b}{2 \rru{3} r}+\frac{m_s m_b^2}{2 r^3 r} \\
\nn && +\frac{(m_b+m_s)}{8 \rru{2} r} + \frac{7 m_b}{8 r' r^2}+\frac{m_b (q^2-m_b^2-m_s^2)}{8 r^2 \rru{2}} \},\\
\Pi_{V_{2B}}&=& 0,\\
\Pi_{V_{3B}}&=& 0,\\
\Pi_{T_{1B}}&=& 0,\\
\Pi_{T_{2B}}&=& -\frac{\qq}{M^2-m^2}\{\frac{m_s(m_b+m_s)}{ r r'}\}+ \frac{m_0^2 \qq}{M^2-m^2}\{ \frac{m_s^3(m_b+m_s)}{2 \rru{3}r}+\frac{m_b^3(m_b+m_s)}{2 r^3 r'}\\
\nn && -\frac{m_s(m_b+m_s)(m_b^2-q^2)}{8 r^2 \rru{2}}+\frac{7 m_b m_s}{8\rru{2}r}-\frac{m_b^2+m_s^2-7m_b m_s}{8 r' r^2}\},\\
\label{nonpertSON}\Pi_{T_{3B}}&=& m_0^2 \qq\{\frac{1}{8 r^2
r'}-\frac{1}{8 r \rru{2}}\}. \eeqa

In the expressions of non-perturbative contributions to correlator
(Eqs. \ref{nonpertILK} to \ref{nonpertSON}), the first terms in
brackets which are proportional to $\qq$ are $d=3$ dimensional, and
the second terms in brackets which are proportional to $m_0^2\qq$
are $d=5$ dimensional contributions corresponding to operators $\qq$
and $\langle \bar{q} \sigma G q \rangle$. The non-perturbative
contributions coming from $d=4$ dimensional operators which are
proportional to $m_d \qq$ and $\langle g^2 G^2 \rangle$ are
neglected.

To obtain the final expression for the sum rules of the form
factors, the quark hadron duality assumption, which states that the
phenomenological and perturbative spectral densities give the same
result when integrated over an appropriate interval, is used. The
quark hadron duality is expressed as\cite{ColangeloBKll}

\beq\label{eq::DUAL} \left[\int_{s_0}^{\infty}
\int_{s'_0}^{\infty}+\int_{s_0}^{\infty}
\int_{0}^{s_0'}+\int_{0}^{s_0} \int_{s_0'}^{\infty}\right] ds ds'
\{\rho^h_{f_i}(s,s',\qsq)-\rho_{f_i}(s,s',\qsq)\}=0,\eeq where $s_0$
and $s'_0$ are the continuum thresholds in $s$ and $s'$ channels,
and $\rho^h(s,s',\qsq)$ is the spectral density of the continuum in
the phenomenological part.

After calculating all spectral densities and nonperturbative
contributions to correlators, by equating the coefficients of the
selected structures from the phenomenological side (Eqs.
\ref{eq:structuresAphen} and \ref{eq::structuresBphen}) and the
theoretical side (Eqs. \ref{eq:structuresA} and
\ref{eq::structuresB}), the QCD sum rules for the form factors
parameterizing $\Mtokab$ matrix elements are found as

\beqa\label{eq::formsfinalA} f_{i,A}(q^2)&=&\frac{m_b+m_d}{f_A m_A
F_B M^2}e^{\frac{M^2}{M_1^2}}e^{\frac{m^2}{M_2^2}}\\
\nn &&\{\frac{-1}{4 \pi}\int_{0}^{s_0} d s \int_{0}^{s'_0} d
s'\Theta
\rho_{f_{i,A}}(s,s',\qsq)e^{\frac{-s}{M_1^2}}e^{\frac{-s'}{M_2^2}}+\Bpi^{nonpert}_{f_{i,A}}\},
\eeqa and \beqa\label{eq::formsfinalB} f_{i,B}(q^2)&=&-i
\frac{m_b+m_d}{f_B (1{\rm GeV}) F_B
M^2}e^{\frac{M^2}{M_1^2}}e^{\frac{m^2}{M_2^2}}\\
\nn &&\{\frac{-1}{4 \pi}\int_{0}^{s_0} d s \int_{0}^{s'_0} d s'
\Theta
\rho_{f_{i,B}}(s,s',\qsq)e^{\frac{-s}{M_1^2}}e^{\frac{-s'}{M_2^2}}+\Bpi^{nonpert}_{f_{i,B}}\}.
\eeqa
where $\Theta\equiv \Theta(1-f(s,s')^2)$ is the unit step function
determining the integration region and $f(s,s')$ is the function
defined in Eq. \ref{eq::inequality}. The expressions for the form
factors of $\toklh$ transitions are obtained by using Eq.
\ref{eq::formmixing}.


\section{Numerical results and discussions}

In this section, the numerical results for the $\tok$ transitions
are presented. The expressions of form factors and the effective
Hamiltonian depend on the parameters $M_1^2$, $M^2_2$, $s_0$,
$s'_0$, on the masses and decay constants of the $K_1$ and $B$
states, on the values of $V_{ij}$, and on the values of the Wilson
coefficients $C_7^{eff}$, $C_9^{eff}$ and $C_{10}$. The values of
the input parameters are presented in table \ref{tb:input}.

\begin{table}[htb]\vskip1cm\caption{\label{tb:input} The values of the input
parameters for numerical analysis.}
\begin{center}
\begin{tabular}{c}
\hline\hline\hskip2cm INPUT PARAMETERS \hskip3cm \\
\hline\hline $M_B=5279$
MeV~~~$\tau_B=(1.525\pm0.002)\times10^{-12}$s.~~~$F_B=0.14\pm
0.01 ~GeV$\cite{16}~~~~~~~~\\
\hline
$m_{s}=95\pm25 ~MeV$~~~~ $m_{b} =(4.7\pm0.07)~GeV$~~~~~ $m_{d}=(3-7)~MeV~$\cite{16}~~~~$\qq\equiv\dd=-(240\pm10 MeV)^3$\cite{loffe}\\
\hline $ m_{K_{1}}(1270)=1.27~GeV$~~~~
$m_{K_{1}}(1400)=1.40~GeV$~~~~\\
$f_{A} =(250\pm13)  ~MeV $~~~~ $f_{B} =(190\pm10) ~MeV
$~~\\
~~$m_{A} =(1.31\pm0.06)  ~GeV $~~~~
$m_{B} =(1.34\pm0.08) ~MeV $ ~\cite{lee,yangana,16}\\
\hline $V_{tb}\mid=0.77^{+0.18}_{-0.24}$~~~~~~~~ $\mid
V_{ts}\mid=(40.6\pm2.7)\times10^{-3}$ \cite{pdg10}~\\
\hline~~~~$C_{10}=-4.669$~~~~~~$C^{eff}_{9}=4.344$~~~~~~$C^{eff}_{7}=-0.313$~~
 \cite{Buras}\\
 \hline $G_F=1.17\times10^{-5}GeV^{-2}$~~~~~~$\alpha=1/129$~~\cite{16}\\
 \hline $M_1^2=16\pm2{\rm GeV}^2$~~~~$M_2^2=6\pm1{\rm
 GeV}^2$~~~~~~$s_0=34\pm4{\rm GeV}^2$~~~~$s'_0=4\pm1{\rm GeV}^2$\\
 \hline\hline
\end{tabular}
\end{center}\vskip1cm
\end{table}

The explicit expressions of the form factors in Eqs.
\ref{eq::formsfinalA} and \ref{eq::formsfinalB} contain four
auxiliary parameters: Borel parameters $M_{1}^2$ and $M_{2}^2$, as
well as the continuum thresholds $s_{0}$ and $s'_{0}$. These are not
physical quantities, hence the physical quantities , form factors,
must be independent of these auxiliary parameters. The working
region of $M_{1}^2$ and $M_{2}^2$ is determined by requiring that
the higher state and continuum contributions are suppressed and the
contribution of the highest order operator must be small. These
conditions are both satisfied in the following regions;
$12~GeV^2\leq M_{1}^{2}\leq20~GeV^2 $ and $4~GeV^2\leq
M_{2}^{2}\leq8~GeV^2 $.  The dependence of form factors $T_{1A}$ and
$T_{1B}$ on Borel masses at $q^2=0$ are plotted in figures
\ref{fig::BorelA}(a) and \ref{fig::BorelA}(b). From the figures it is
found that the results are stable in the working region of Borel
mass parameters.

The continuum thresholds $s_{0}$ and $s_{0}' $ are determined by
two-point QCD sum rules and related to the energy of the excited
states. The form factors which are the physical quantities defining
the transitions, should be stable with respect to the small
variations of these parameters. In general, the continuum thresholds
are taken to be around $(m_{hadron}+0.5)^2$ \cite{Aliev1,
Colangelo1, Shifman1}. The dependence of form factors $T_{1A}$ and
$T_{1B}$ on continuum thresholds at $q^2=0$ are plotted in figures
\ref{fig::BorelA}(c) and \ref{fig::BorelA}(d). From the figures it is
found that the results are stable for variations of $s_0$ and
$s_0'$.

The sum rules expressions for the form factors are truncated at $7
~GeV^2$. In order to extend our results to the whole physical
region, i.e., $0\leq q^{2}<(m_{B}-m_{K_{1}})^{2}$ and for the
reliability of the sum rules in the full physical region, a fit
parametrization is applied such that in the region $-10 GeV^{2}\leq
q^{2} \leq -2~GeV^{2}$, where the spectral integrals can be handled
safely by applying Cutkovsky rules. This parametrization coincides
with the sum rules predictions. To find the extrapolation of the
form factors in the whole physical region, the fit function is
chosen as

 \begin{equation}\label{17au}
 f_{i}(q^2)=\frac{f_i(0)}{1-a \hat{q}+b \hat{q}^2}.
\end{equation}

The values for a, b and $f_i(0)$ are given in Table \ref{tb::fit1}
and \ref{tb::fit2} for the form factors of $\Mtoka$ and $\Mtokb$
matrix elements respectively. The errors in the values of $f_i(0)$
in tables \ref{tb::fit1} and \ref{tb::fit2} are due to uncertainties
in sum rule calculations and also due to errors in input parameters.

\begin{table}[h]\vskip0.1cm\caption{\label{tb::fit1} The fit parameters and coupling constants for  $\Mtoka$ matrix elements.}
\begin{center}
\begin{tabular}{cccc}
\hline\hline \hskip1cm $f_i$\hskip1cm&$f_i(0)$\hskip2cm&$a$\hskip2cm&$b$ \hskip3cm\\
\hline\hline \hskip1cm $\faa$& $0.47\pm0.08$&$0.74$&$-0.41$ \\
\hline  \hskip1cm $\fvba$& ~~$0.35\pm0.07$~~&~~$0.52$~~&~~$-1.2$~~\\
\hline  \hskip1cm $\fvia$& $0.36\pm0.07$&$0.41$&$-0.74$\\
\hline  \hskip1cm $\fvua$& $-(0.39\pm0.08)$&$0.45$&$-0.27$\\
\hline  \hskip1cm $\ftba$& $0.38\pm0.08$&$0.73$&$-0.36$\\
\hline  \hskip1cm $\ftia$& $0.38\pm0.09$&$0.67$&$-0.26$\\
\hline  \hskip1cm $\ftua$& $0.36\pm0.07$&$0.42$&$-0.15$\\
 \hline\hline
 \end{tabular}
\end{center}\vskip0.1cm
\end{table}

\begin{table}[h]\vskip0.1cm\caption{\label{tb::fit2} The fit parameters and coupling constants for $\Mtokb$ matrix elements.}
\begin{center}
\begin{tabular}{cccc}
\hline\hline  \hskip1cm$f_i$\hskip1cm&$f_i(0)$\hskip1cm&$a$\hskip2cm&$b$ \hskip3cm\\
\hline\hline \hskip1cm$\fab$& $-0.31\pm0.06$&$1.3$&$0.37$\\
\hline  \hskip1cm$\fvbb$& $-0.40\pm0.08$&$1.4$&$-0.10$\\
\hline  \hskip1cm$\fvib$& $-0.34\pm0.06$&$1.2$&$0.37$\\
\hline  \hskip1cm$\fvub$&~~ $0.39\pm0.08$~~&~~$1.1$~~&~~$0.46$~~\\
\hline  \hskip1cm$\ftbb$& $-0.22\pm0.05$&$1.1$&$0.24$\\
\hline  \hskip1cm$\ftib$& $-0.21\pm0.07$&$1.3$&$0.081$\\
\hline  \hskip1cm$\ftub$& $-0.26\pm0.04$&$1.1$&$0.27$\\
 \hline\hline
\end{tabular}
\end{center}\vskip0.1cm
\end{table}

The $q^2$ dependence of $f_{i,A}$ and $f_{i,B}$, the sum rules
predictions and also the fit results, are plotted in the range
$-10\leq q^2\leq M^2-m^2$ in figures \ref{fig::k1a} and
\ref{fig::k1b}. It is seen from tables \ref{tb::fit1} and
\ref{tb::fit2}, and from figures \ref{fig::k1a} and \ref{fig::k1b}
that the form factors of $\toka$ transition, i.e. $f_{i,A}$, and the
form factors of $\tokb$ transition, i.e. $f_{i,B}$ are opposite in
sign.


For the transitions to physical states, i.e. for $\toklh$
transitions, the dependence of the form factors of $\tokl$ on the
mixing angle $\thetaK$ are plotted in figure \ref{fig::vl}, and the
dependence of form factors of $\tokh$ on the mixing angle $\thetaK$
are plotted in figure \ref{fig::tl} at $q^2=0$. The region between
two black dashed vertical lines is the region estimated as
$\thetaK=(-34\pm 13)\degr$\cite{hatanakayang}. It is seen from
figures \ref{fig::vl}(a) and \ref{fig::vl}(b) that the absolute
values the form factors of $\tokl$ transition are maximum  at
$\thetaK=-(45\pm5)\degr$, and their values are zero at
$\thetaK=42\pm5\degr$. For the form factors of $\tokh$ transitions,
it is seen from figures \ref{fig::tl}(a) and \ref{fig::tl}(b) that
the absolute values of the form factors are maximum at
$\thetaK=40\pm5\degr$, their values are zero at
$\thetaK=-(47\pm7)\degr$. Since the region $\thetaK=-(47\pm7)\degr$
in which form factors are zero coincides with the region
$\thetaK=(-34\pm 13)\degr$, to obtain a precise prediction of the
form factors, the mixing angle should be determined more precisely.

Finally, the branching fractions to leptonic final states $\epm$,
$\mupm$ and $\taupm$ for $\thetaK=-34\degr$ are also estimated by
integrating the partial width in Eq. \ref{eq::partialwidth}. The
results are presented in table \ref{tb::BR} in comparison with the
results found in \cite{hatanakayang}. The first errors in our
results are due to uncertainties from sum rule calculations and
input parameters, and the second errors are due to uncertainty in
the mixing angle $\thetaK$. Our results are in good agreement with
the results found in \cite{hatanakayang}.

\begin{table}[h]\vskip1cm\caption{\label{tb::BR} The branching fractions of $\toklh$ decays for $\thetaK=-34\degr$ .}
\begin{center}
\begin{tabular}{ccc}
\hline\hline mode\hskip5cm&this work\hskip5cm&\hskip3cm\cite{hatanakayang}\hskip5cm\\
\hline\hline${\cal B}(\kkkl\epm)$&$(2.11\pm0.82\err{0.42}{0.52})\ttn{-6}$&$(2.5\err{1.4+0.0}{1.1-0.3})\ttn{-6}$\hskip3cm\\
\hline${\cal B}(\kkkl\mupm)$&$(2.10\pm0.81\err{0.41}{0.49})\ttn{-6}$&$(2.1\err{1.2+0.0}{0.9-0.2})\ttn{-6}$\\
\hline${\cal B}(\kkkl\taupm)$&$(0.42\pm0.21\err{0.11}{0.15})\ttn{-7}$&$(0.8\err{0.4+0.0}{0.3-0.1})\ttn{-7}$\\
\hline\hline${\cal B}(\kkkh\epm)$&$(1.1\pm0.4\err{0.4}{0.5})\ttn{-7}$&$(0.9\err{0.3+2.3}{0.3-0.4})\ttn{-7}$\\
\hline${\cal B}(\kkkh\mupm)$&$(1.0\pm0.4\err{0.4}{0.5})\ttn{-7}$&$(0.6\err{0.2+1.8}{0.1-0.2})\ttn{-7}$\\
\hline${\cal B}(\kkkh\taupm)$&$(0.3\pm0.2\err{0.1}{0.1})\ttn{-8}$&$(0.1\err{0.0+0.5}{0.0-0.1})\ttn{-8}$\\
\hline\hline
\end{tabular}
\end{center}\vskip1cm
\end{table}

The experimental bounds on $\toklh$ decays can be obtained from the
inclusive $B \to X_s \lpm$ decays. The current averages on inclusive
decays are\cite{pdg10,HFAG}:

\begin{eqnarray}\label{HFAG}
\nn {\cal B}(B\to X_s \epm)&=&(4.7\pm1.3)\ttn{-6}[(4.91\err{1.04}{1.06})\ttn{-6}],\\
\nn{\cal B}(B\to X_s\mupm)&=&(4.3\pm 1.2)\ttn{-6}[(2.23\err{0.97}{0.98})\ttn{-6}],\\
{\cal B}(B\to X_s \lpm)&=&(4.5\pm1.0)\ttn{-6}[(3.66\err{0.76}{0.77})\ttn{-6}],
\end{eqnarray}
where the first averages are from PDG\cite{pdg10}, and the second values in square brackets are the recent HFAG averages\cite{HFAG}.

The results found in this work (table \ref{tb::BR}) are consistent with these average values except $B \to K_1 (1270)\mu^+\mu^-$. For $b \to s\mu^+\mu^-$
channel, when the measured branching fractions of exclusive decays $B\to K\mu^+\mu-$ and $B\to K^* \mu^+\mu^-$   which amount $(1.63\pm0.21)\times 10^{-6}$\cite{pdg10} are considered, the room left for  other exclusive decays including $B \to K_1 (1270)\mu^+\mu^-$ is about $(0.6 \sim 2.6)\times 10^{-6}$. However when the errors are considered, even this channel is still consistent with the data.
And also when the dependence of branching ratios on $\thetaK$ are considered(fig. \ref{fig::brs}), it can be seen that the value of the
branching ratio of the $B \to K_1 (1270)\mu^+\mu^-$ can be smaller depending on the mixing angle $\thetaK$ .

In conclusion, we have calculated the form factors of $\tokab$
transitions using three point QCD sum rules approach. We analyzed
the $q^2$ behaviors of the form factors of $\tokab$ transitions.
Considering the axial vector mixing angle $\thetaK$, we estimated
the form factors of $\toklh$ transitions, i.e. transitions into
physical states and analyzed their dependence on the mixing angle
$\thetaK$ at $q^2=0$. Using these results we estimated the branching
fractions into final leptonic states. We conclude that the
transitions $\toklh$ can be observed at LHC and further B factories
and measurements on the mixing angle $\thetaK$ can be performed.

%
\begin{acknowledgments}
The authors would like to thank to T. M. Aliev, H. D. thanks
to G. Erkol, J. Smith and K. Azizi, A.{\" O}. thanks to K-C. Yang
for useful discussions during this work. The work of A.{\" O}. have
been been supported in part by the European Union (HadronPhysics2
project
 Study of strongly interacting matter.).
\end{acknowledgments}


\newpage
\appendix
\section{Figures}

\begin{center}
\begin{figure}[h]\vskip2cm
\begin{center}
\includegraphics[width=7cm]{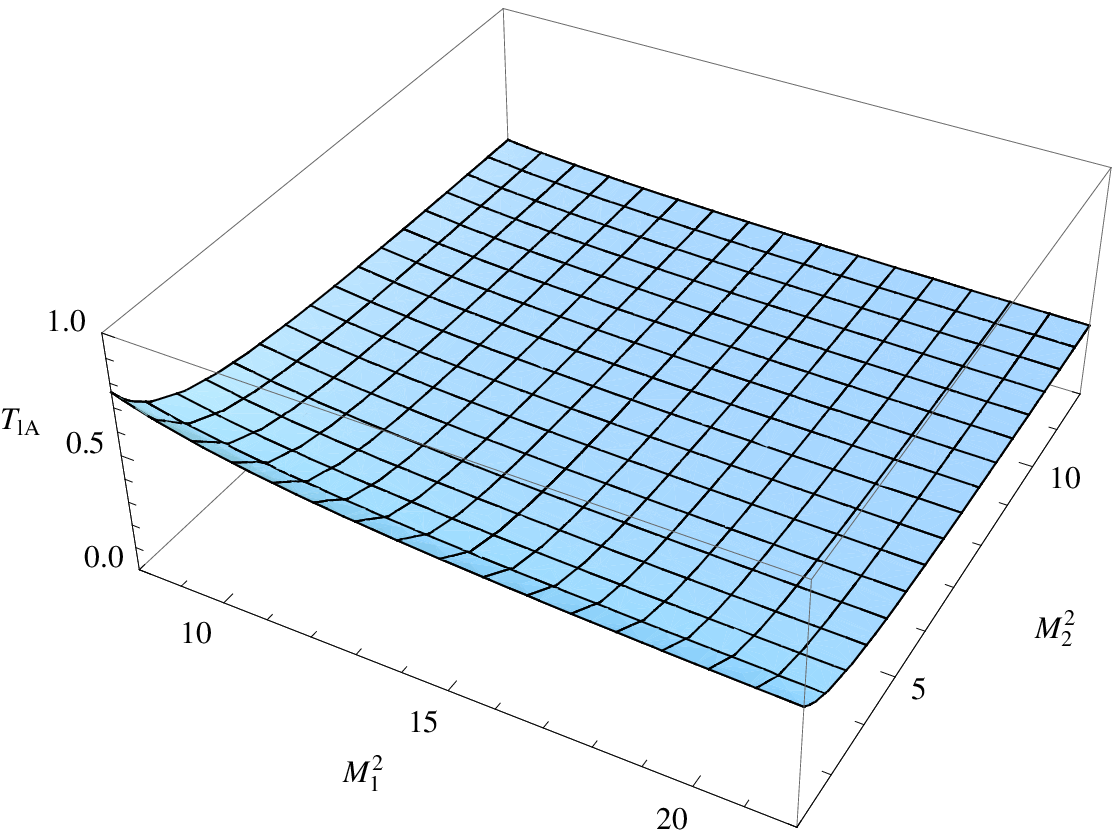}(a)~~~~~~~\includegraphics[width=7cm]{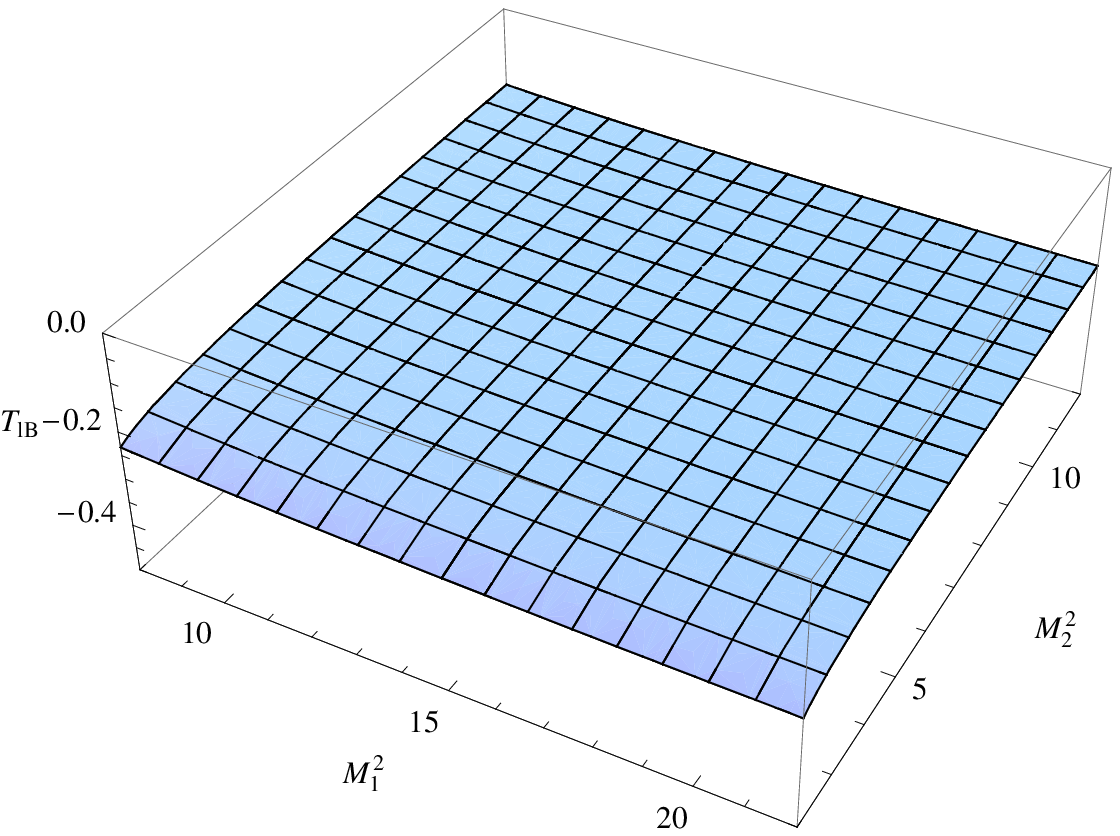}(b)
 \vskip2mm
\includegraphics[width=7cm]{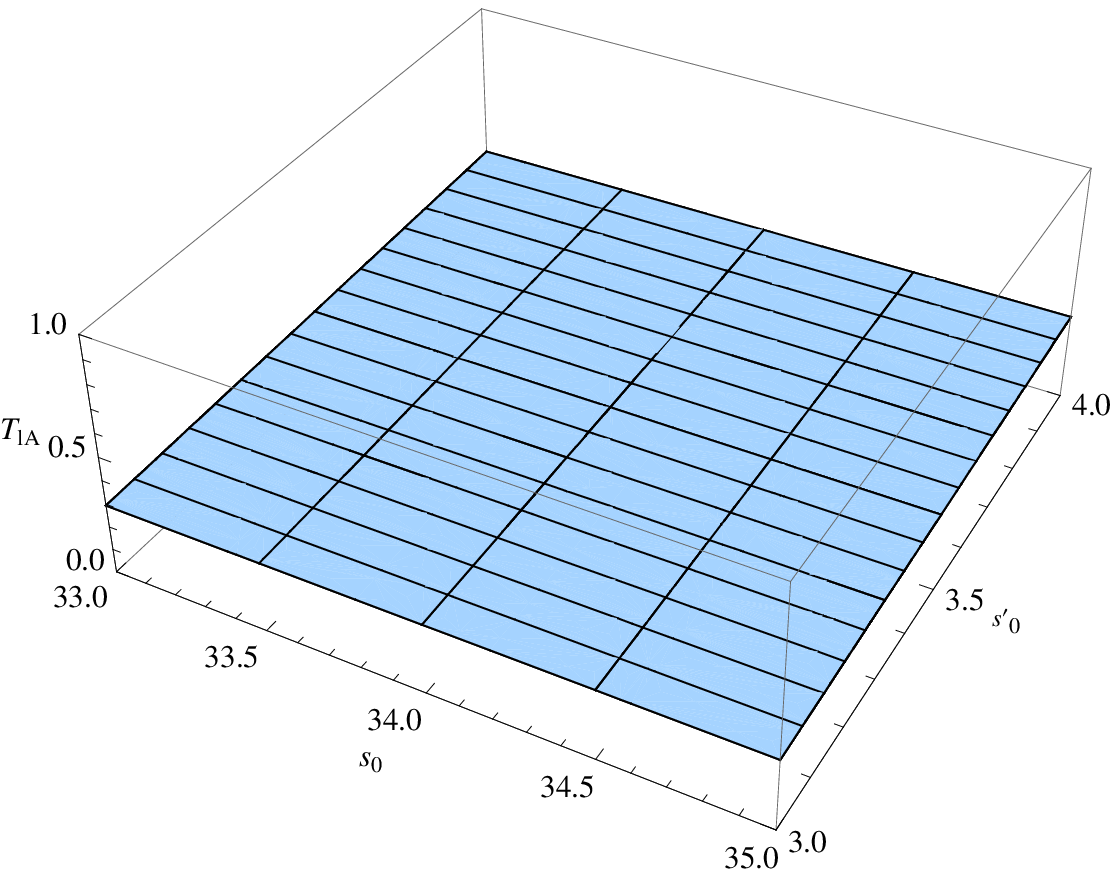}(c)~~~~~~~\includegraphics[width=7cm]{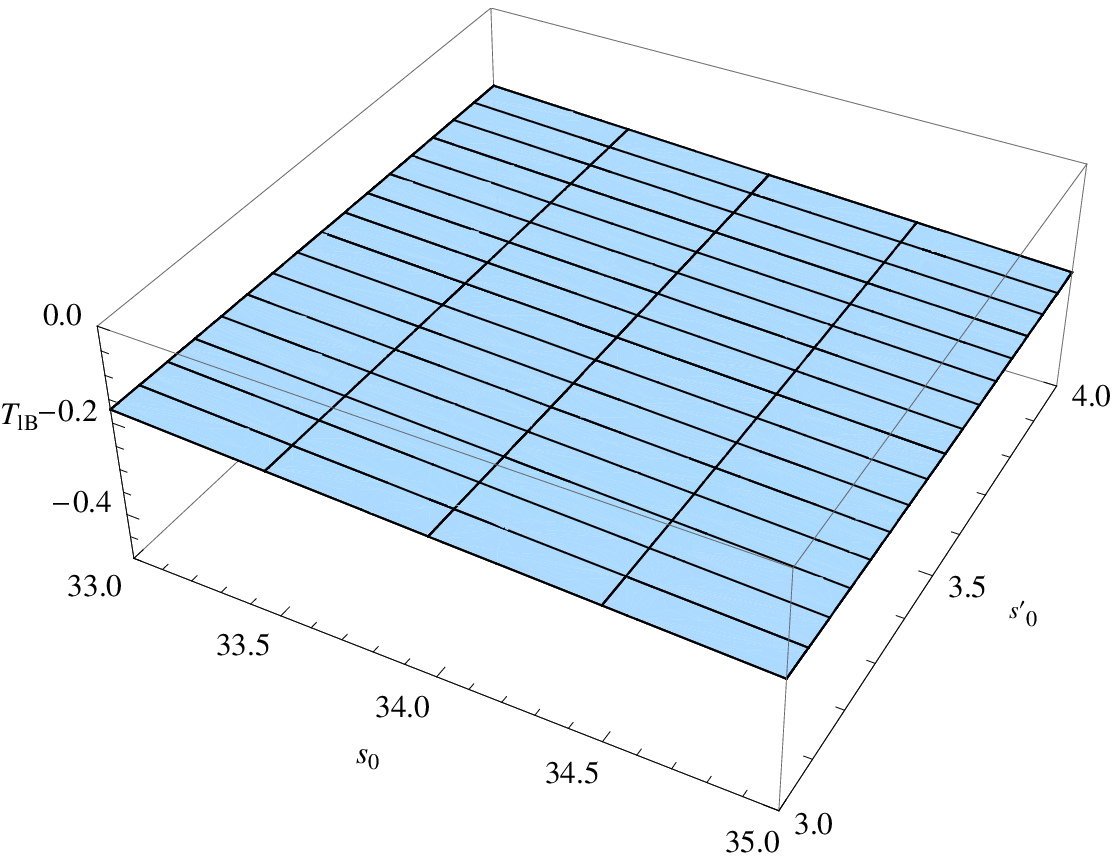}(d)

\caption{\label{fig::BorelA}(a,b):The dependence of the form factors
$\ftba$(a) and $\ftbb$(b) on Borel mass parameters $M_1^2$ and
$M_2^2$ at $q^2=0$ for $s_0=34 GeV^2$ and $s_0'=4 GeV^2$.(c,d): The
dependence of the form factors $\ftba$(c) and  $\ftbb$(d) on
continuum thresholds $s_0$ and $s'_0$ at $q^2=0$ for $M_1^2=16
GeV^2$ and $M_2^2=6 GeV^2$ .}
\end{center}
\end{figure}
%


%



\begin{figure}[htb]
\begin{center}
\includegraphics[width=7.5cm]{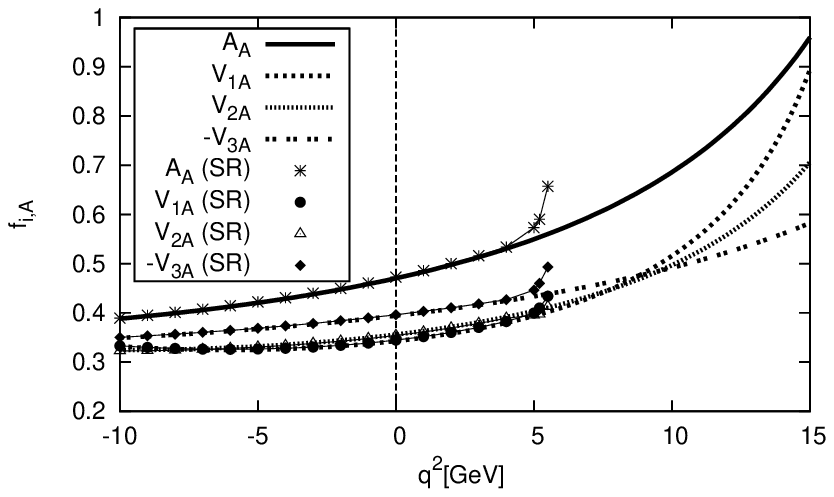}(a)~~~~~~~~\includegraphics[width=7.5cm]{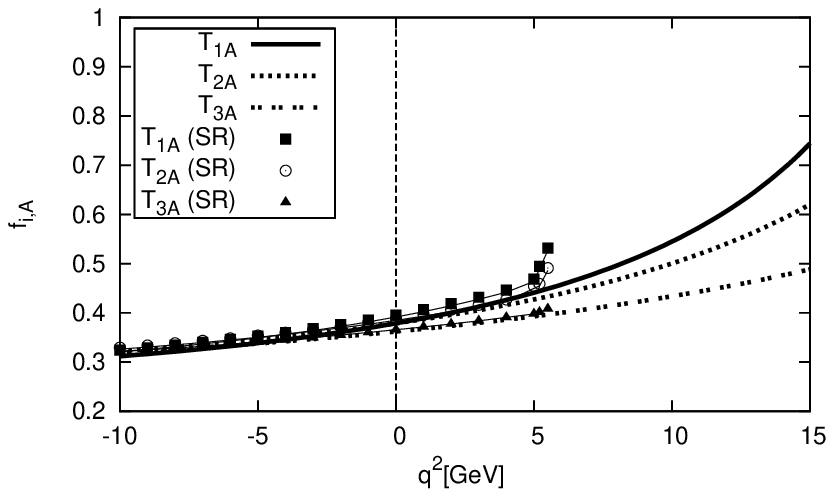}(b)
 \vskip2mm \caption{\label{fig::k1a}The $\qsq$ dependence of the vector form factors(a) and tensor form factors(b) of $\Mtoka$ matrix element. The sum rules predictions for the form factors also shown with data points connected with a thin line. }
\end{center}
\end{figure}

\begin{figure}[tbh]
\begin{center}
\includegraphics[width=7.5cm]{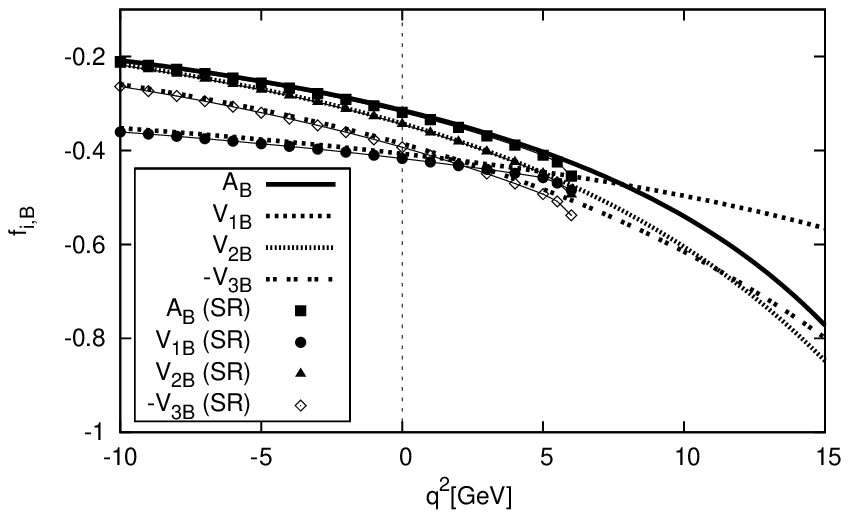}(a)~~~~~~~~\includegraphics[width=7.5cm]{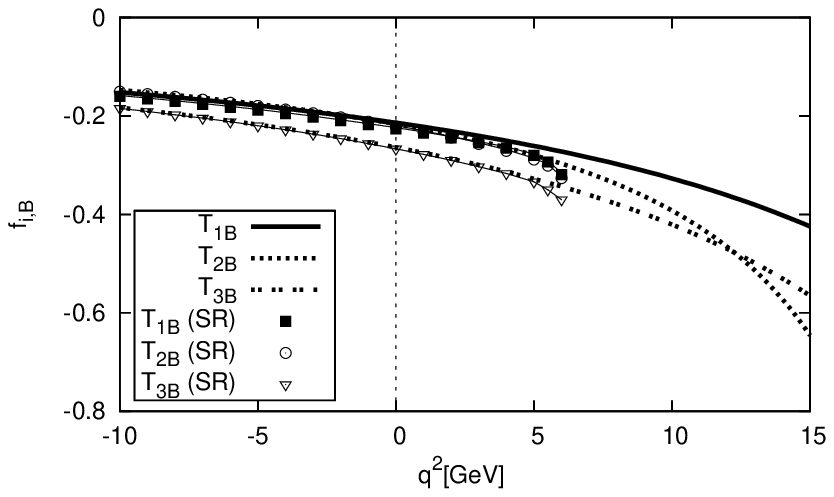}(b)
 \vskip2mm \caption{\label{fig::k1b}The $\qsq$ dependence of the vector form factors(a) and tensor form factors(b) of $\Mtokb$ matrix element. The sum rules predictions for the form factors also shown with data points connected with a thin line.}
\end{center}
\end{figure}


%
\begin{figure}[htb]
\begin{center}
\includegraphics[width=7.5cm]{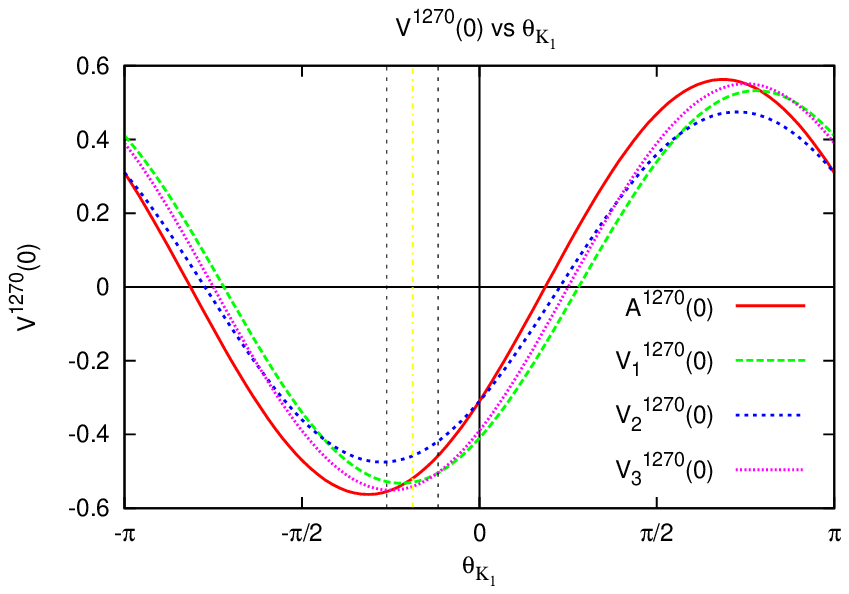}(a)~~~~~~~~\includegraphics[width=7.5cm]{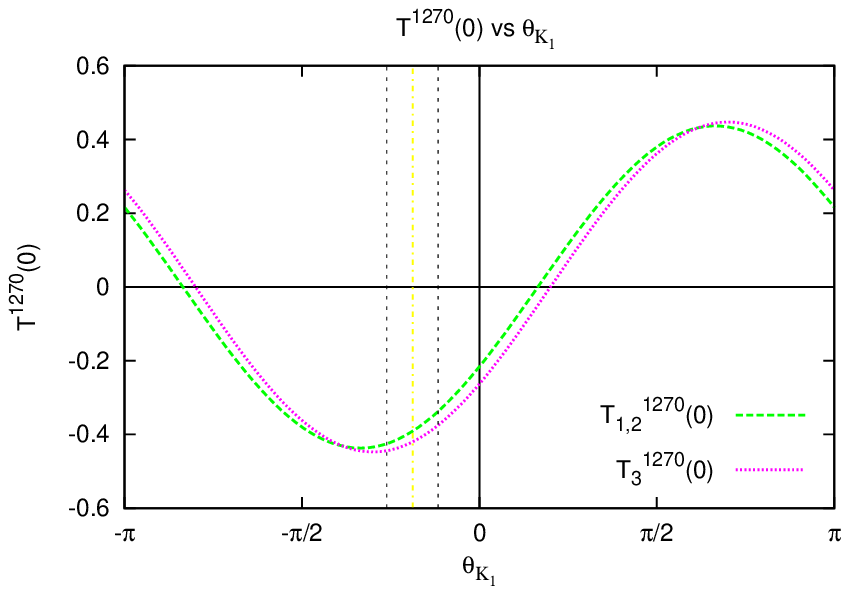}(b)
 \vskip2mm \caption{\label{fig::vl}The $\thetaK$ dependence of the vector form factors(a) and tensor form factors(b) of $\tokl$ at $\qsq=0$.}
\end{center}
\end{figure}
%
%
\begin{figure}[htb]
\begin{center}
\includegraphics[width=7.5cm]{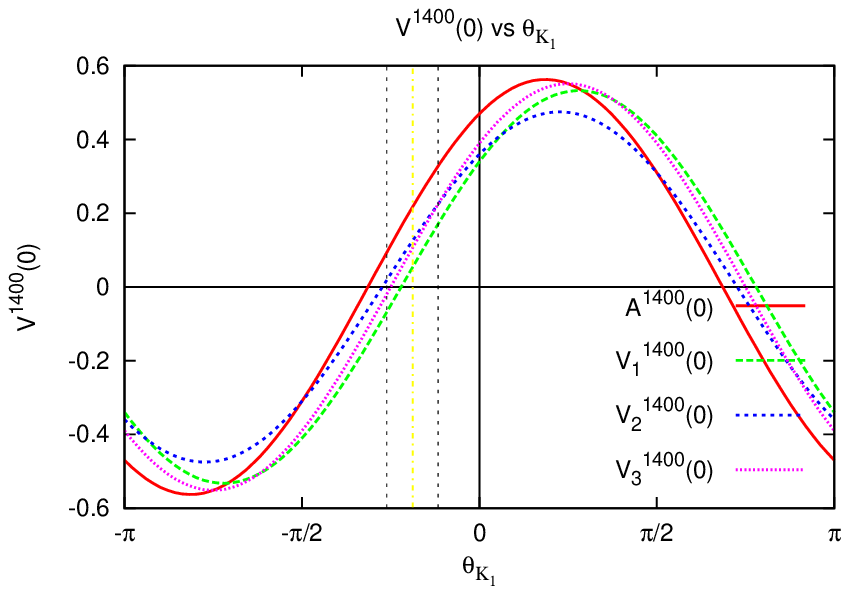}(a)~~~~~~~~\includegraphics[width=7.5cm]{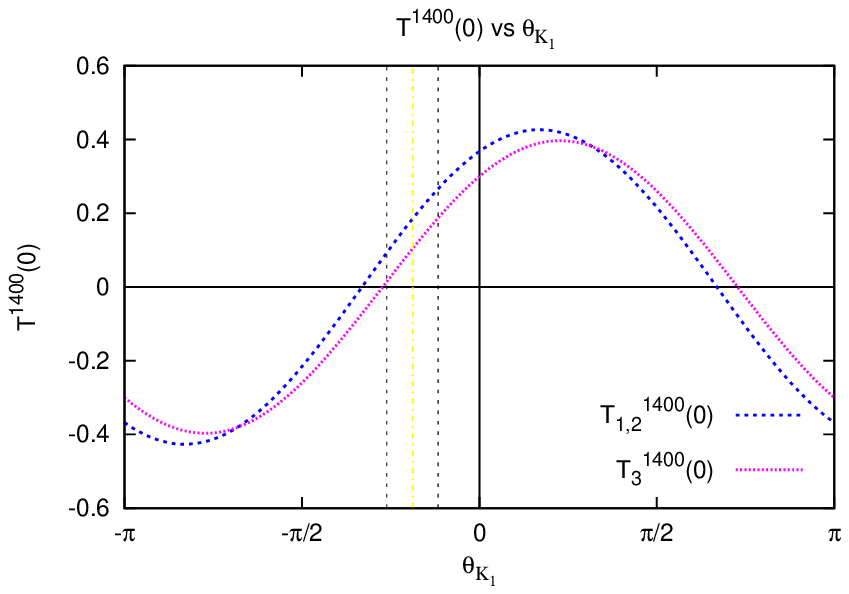}(b)
 \vskip2mm \caption{\label{fig::tl}The $\thetaK$ dependence of the vector form factors(a) and tensor form factors(b) of $\tokl$ at $\qsq=0$.}
\end{center}
\end{figure}
%

\begin{figure}[htb]\vskip2cm
\begin{center}
\includegraphics[width=11cm]{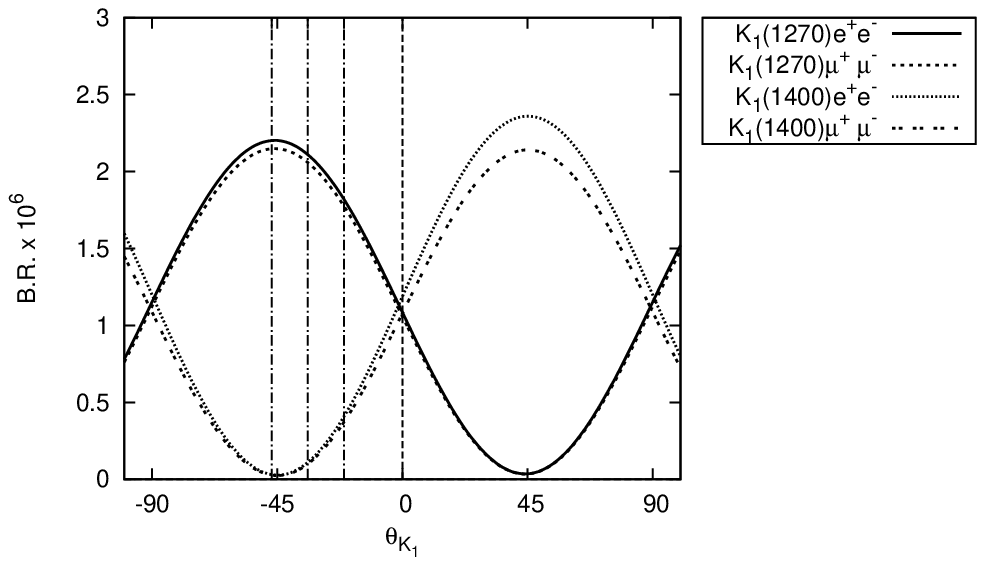}
 \vskip2mm \caption{\label{fig::brs}The $\thetaK$ dependence of the branching ratios of $B\to \kkkl \epm$(solid), $B\to \kkkl \mupm$(dashed), $B\to \kkkh \epm$(dotted) and $B\to \kkkl \mupm$(double dashed) channels. }
\end{center}
\end{figure}
\clearpage
\end{center}

\end{document}